\documentclass[floatfix,twocolumn,aps,prl,amsmath,amssymb]{revtex4}
\usepackage[dvips]{graphicx}
\usepackage{color}
\usepackage[colorlinks=true ,citecolor=blue]{hyperref}

\begin{document}

\title{ Orbital Order, Superconductivity, Pseudogap, Spectral Weight and Phase Diagrams in High-Tc Cuprates 

  }
\author{E. C. Marino}\thanks{marino@if.ufrj.br}
\affiliation{Instituto de F\' isica, Universidade Federal do Rio de Janeiro, C.P. 68528, Rio de Janeiro, RJ, 21941-972, Brazil.}%

\date{\today}

\begin{abstract}

After providing a brief genealogy of our recently proposed model for High-Tc cuprates, we investigate a few pending details of our model, which remained to be explained.  The first issue concerns the microscopic mechanism that produces an attractive interaction between neighboring holes. We show that a peculiar arrangement of the $p_x$ and $p_y$ oxygen orbitals makes the mutual magnetic interaction of the holes with the localized copper ions, mediated by the ferromagnetic fluctuations of the localized copper spins, to produce a net attractive interaction between themselves, which is responsible for the emergence of a superconducting phase. We also study the connection existing between the proposed pseudogap order parameter and the spectral density. We show that the occurrence of two sharp peaks in the latter, between which the density of states suffers a depletion is a direct consequence of the d-wave character of the pseudogap order parameter dependence on $\mathbf{k}$, which breaks the 90$^\circ$-rotation symmetry of the oxygen lattices. The peak separation in the spectral density works effectively as an overall pseudogap order parameter for the cuprates. We explicitly calculate  the spectral density in the strange metal and pseudogap phases of Bi2212, at different temperatures, and show that our results compare very well with the experimental data.  We also derive a new relation among the parameters determining the phase diagrams $T_c(x)\times x$ and $T^*(x)\times x$, for the case of multilayered cuprates and revisit the obtainment of these in comparison to the experimental data.
We, finally, examine the effect of an external applied pressure on $T_c(x)$ on the light of this new relation for multi-layered cuprates.

\end{abstract}

\maketitle

{\bf 1) Introduction}\\
\bigskip

 We have developed, recently, a comprehensive theory for the superconductivity in High-Tc copper oxides \cite{M1, M2, M3}. This has been applied in the description of the superconducting (SC) phase, as well as of most of the normal phases of the cuprates, such as the pseudogap (PG), strange metal (SM), Fermi liquid (FL) and the crossover between the last two. The theory also possesses a magnetic sector, which is expressed in terms of the nonlinear sigma model and allows for the obtainment of the N\' eel temperature as a function of doping \cite{ecmmbsn}, thus delimiting the magnetically ordered antiferromagnetic (AF) phase.
 
One of the nice features of our theory is the fact that it can be tested, namely, we can calculate, out of it, the values of physical quantities which can be compared with the experimental data available for several cuprate compounds. 
Indeed, based on this theory, we have been able to obtain analytical expressions for the SC and PG transition temperatures as a function of doping, namely, $T_c(x)$ and $T^*(x)$; the dependence of $T_c$ on the number of $CuO_2$ planes, as well as on an applied pressure $T_c(x,P)$. We have also established the pressure independence of the PG temperature $T^*(x)$. Furthermore, starting from such theory, we obtained analytical expressions for the resistivity as a function of the temperature $\rho(T)$ in the PG, SM and FL phases and also in the crossover between the last two. Finally, we have derived expressions for the magnetoresistivity in the overdoped regime. All the above mentioned results are in excellent agreement with the experimental data for LSCO, YBCO and the Bi, Hg and Tl families of cuprates.

In the present study, we start by  presenting a brief genealogy of our theory, that will serve to understand how it has evolved out of well-known first principle models and also to locate its position in the scenario of theories for High-Tc superconductivity in cuprates.

We, then, proceed to examine several key issues of High-Tc cuprates that were still pending in the solution provided by our theory, as listed below.

{\bf 1) Superconductivity: Interaction.}  We explore the most basic features underlying the mechanism responsible for the occurrence of superconductivity in cuprates. We show, in particular, how this mechanism is closely related to the symmetry breakdown that produces a $d$-wave symmetric SC gap. Indeed, we demonstrate that when the oxygen 
$p_x$ and $p_y$ orbitals 
organize themselves in such a way as to break the lattice symmetry under rotations of 90$^\circ$, the magnetic interaction of neighboring holes with the closest $Cu^{++}$ ion, mediated by the ferromagnetic fluctuations of such localized copper ion, produces an effective {\it attractive} interaction between these holes. This is the interaction responsible for superconductivity in the High-Tc cuprates.

This interaction resembles the Kondo interaction, in the sense that it involves a magnetic interaction between localized and itinerant spins. The crucial difference, however, resides in the fact that in the Kondo system, the itinerant spins are essentially free conduction band electrons, whereas, in the case of cuprates, conversely, the itinerant spins are holes belonging to a sophisticated structure of $p_x$ and $p_y$ orbitals organized on a bipartite square lattice in a particular way such that the mutual hybridization of neighboring holes with the $Cu^{++}$ ion $d_{x^2-y^2}$ orbital, produces the attractive interaction responsible for Cooper pair formation and for the superconductivity in cuprates. The mediation of this SC interaction is made by the {\it ferromagnetic} fluctuations of the localized copper spins (see Fig. 2). 

{\bf 2) Pseudogap: Interaction.}
A remarkable feature of the physics of cuprates is the presence of the pseudogap phase \cite{htsc4,77}, which in our theory is associated to an order parameter $M(k_x,k_y)$ given by the ground-state expectation value of the exciton creation operator. 
Coulomb repulsion between holes is the interaction that competes with the previous SC interaction. It leads to exciton (electron-hole) pairs formation while the former leads to Cooper pair formation. The condensation of each, respectively, leads to the SC and PG phases.

{\bf 3) Pseudogap: Order Parameter.}
What about the PG order parameter?
One frequently hears in the community the question/remark: is there an order parameter, which is non-vanishing in the whole PG phase and therefore unequivocally characterizes it? We will demonstrate here that this question has a positive answer: the PG order parameter is proportional to the energy separation between the two peaks that form in the spectral weight in the PG region. As we shall see, the depletion of energy states in between the two peaks is a direct consequence of the $k_x,k_y$-dependence of the PG order parameter.

Starting from the dispersion relation of our model, we calculate the spectral weight, both in the SM and PG phases, exhibiting the sharp difference that exists between both. We compare our results for the spectral weight at different temperatures in the PG phase, with the experimental measurements made in Bi2212 \cite{sw} and show that they are in good agreement.

{\bf 4) The SC and PG Phase Diagrams}
One of the nice features of our model is the obtainment of the SC and PG phase diagram of the cuprates \cite{M1}. In the present study, we re-derive these phase diagrams for the case of multi-layered cuprates, taking into account the relation that exists among the $\gamma$, $x_0$ and $\eta$ parameters \cite{M1,M2,M3} when, $N$, the number of $CuO_2$ planes per unit cell, is larger than one (see Sect. 5.3 for a derivation of such new relation).

{\bf 5) Pressure Effects on Tc.}
Finally, using the relation derived in Sect. 5.3, we obtain very accurate expressions for the influence of an external pressure on $T_c(x)$, by adjusting a single parameter in each compound. We compare our theoretical result with experimental data for Hg1212 and Hg1223. 
\\
\bigskip

\vfill
\eject

{\bf 2)The Genealogy of the Effective Hamiltonian }\\
\bigskip

We are going to examine here the genealogy of the Hamiltonian we use in the theory for superconductivity (SC) of high-Tc cuprates, which we introduced recently in \cite{M1} and subsequently explored in \cite{M2,M3}.
This will help to build a perspective on the situation of our model and trace its origin back to the first principles.
\\
\bigskip

{\bf 2.1) The First Generation  Hamiltonian}\\

\bigskip

There are clearly three main actors playing a central role in the physics of high-Tc cuprates drama, which unfolds on the stage of the $CuO_2$-planes: the $3d$ electrons of the $Cu^{++}$ ions, and the $2p_x$ and $2p_y$ electrons of the $O^{--}$ ions. Let us call $d^\dagger_{\sigma}(\mathcal{R}_I)$, $p^\dagger_{x,\sigma}(\textbf{R}_i)$ and $p^\dagger_{y,\sigma}(\textbf{R}_i+\textbf{d}_j)$the creation operators of electrons with spin $\sigma=\uparrow,\downarrow$ belonging to these orbitals.  It is quite plausible that the essential features of the behavior of these electrons should be properly described  by the so-called Three Bands Hubbard Model (3BHM) \cite{3bhm,3bhm1,thesis_3band,num_3band_1,num_3band_2,num_3band_3}, which corresponds to the Hamiltonian

\begin{eqnarray}& &
\hspace{-5mm}H_{3BH} =-t_{pp} \sum_{\textbf{R},\textbf{d}_i}\sum_{\sigma}p_{x}^\dagger(\textbf{R},\sigma)p_{y}(\textbf{R}+\textbf{d}_i,\sigma) +hc
\nonumber \\
&\ &
\hspace{-5mm}-t_{pd} \sum_{\mathcal{R}}\sum_{\sigma}d^\dagger(\mathcal{R},\sigma)\left[ p_{x}(\mathcal{R}+\frac{a}{2}\hat{x},\sigma)+ p_{x}(\mathcal{R}-\frac{a}{2}\hat{x},\sigma) +\right .
\nonumber \\
&\ &
\left .
p_{y}(\mathcal{R}+\frac{a}{2}\hat{y},\sigma)+p_{y}(\mathcal{R}-\frac{a}{2}\hat{y},\sigma)\right]+hc
\nonumber \\
&\ &
U_p \sum_{\textbf{R}} n^{p_x}_\uparrow n^{p_x}_\downarrow + U_p \sum_{\textbf{R}+\textbf{d}} n^{p_y}_\uparrow n^{p_y}_\downarrow + U_d \sum_{\mathcal{R}} n^{d}_\uparrow n^{d}_\downarrow 
\nonumber \\
& &
U_{pd} \sum_{\mathcal{R},\textbf{R}} n^{d}_\uparrow n^{p_x}_\downarrow + U_{pd} \sum_{\mathcal{R},\textbf{R}+\textbf{d}} n^{d}_\uparrow n^{p_y}_\downarrow 
\nonumber \\
& &
\label{0a}
\end{eqnarray}
where $\mathcal{R}$ represents the copper ions lattice positions, $\textbf{R}$ denotes the positions of the oxygen atoms in one sublattice and $\textbf{d}_i$, i=1,...,4 its four nearest neighbors located in the opposite sublattice.
This the First Generation Model.\\
\bigskip

{\bf 2.2) The Second Generation Effective Hamiltonian}\\

As we dope the system by the introduction of holes, these become the main actors on the stage of $CuO_2$-planes. 

There are two main branches in this genealogy tree, each one containing
a model Hamiltonian, which describes the holes: the t-J Model \cite{zhangrice} and its associated RVB solution \cite{RVB1,RVB2,RVB3}, on one side \cite{cti1,cti2,cti3}and
the Spin-Fermion (SF) model, which is a natural evolution of the 3BHM \cite{sf} and comprises the three terms below \cite{M1}: the hopping term, the AF Heisenberg term and the Kondo-like term, namely
\bigskip
\begin{eqnarray}& &
\hspace{-5mm}H_0 =-t_{p} \sum_{\textbf{R},\textbf{d}_i}\sum_{\sigma}\psi_{A,\sigma}^\dagger(\textbf{R})\psi_{B,\sigma}(\textbf{R}+\textbf{d}_i) +hc
\nonumber \\
& &
\hspace{-5mm}H_{AF} = J_{AF}\sum_{\langle IJ \rangle} \textbf{S}_I\cdot  \textbf{S}_J
\nonumber \\
& &
\hspace{-5mm}
H_K =J_K\sum_{I} \textbf{S}_I\cdot\left[\sum_{\textbf{R}\in I} \eta_A \eta_C\  \mathcal{S}_A +
\sum_{\textbf{R}+\textbf{d}\in I} \eta_B \eta_C' \    \mathcal{S}_B   \right ]
\nonumber \\
& &
\label{0a}
\end{eqnarray}
where
$\psi^\dagger_A(\textbf{R})$, $\psi^\dagger_B(\textbf{R}+\textbf{d})$ are the hole creation operators on sites $\textbf{R}$ and $\textbf{R}+\textbf{d}$, respectively, of the $A,B$ oxygen sub-lattices. 

In the equation above,
\begin{equation}
\mathcal{S}_{A,B}=\frac{1}{2}\psi^\dagger_{(A,B)\alpha} \vec{\sigma}_{\alpha\beta}\psi_{(A,B)\beta} \ \ 
\label{1b}
\end{equation}
 which is the spin operator of holes belonging to the $p_x,p_y$ oxygen orbitals, which are associated, respectively, to the  $A$ and $B$ sub-lattices. The sign factors $\eta_A,\eta_B=\pm 1$ originate in the exchange integrals involving the atomic orbitals and are determined by the sign of the half-portion of the $p_x$ and $p_y$
oxygen orbitals that overlaps with the one-fourth portion of the $d_{x^2-y^2}$ copper orbitals, whose sign we denote by either $\eta_{C}=\pm 1$ or $\eta_{C}'=\pm 1$. The sign factors $\eta_A,\eta_B,\eta_C,\eta_C'$ are explicitly displayed in the figures below.

The Hamiltonian of our theory for the cuprates descends from the second branch, namely, the SF model. 
In what follows, we focus on two possible oxygen orbital sign configurations, which are represented, respectively in red and green. 
\begin{figure}
	[h]
	\centerline{
		\includegraphics[scale=0.5]{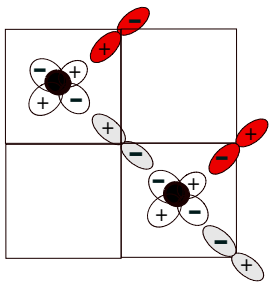}}
		\caption{The base of a crystal structure of the oxygen orbitals around the $Cu^{++}$ ions in the $CuO_2$ planar lattice, which minimizes the energy in the SC phase.  This  configuration of oxygen orbitals leads the magnetic interaction between localized and itinerant spins to a hole-attractive interaction. The black circles represent Cu ions, whereas the oxygen ions belonging to sub-lattices $A$ and $B$ are represented in red and gray, respectively.}
	\label{f3}
\end{figure}
\begin{figure}
	[h]
	\centerline{
		\includegraphics[scale=0.5]{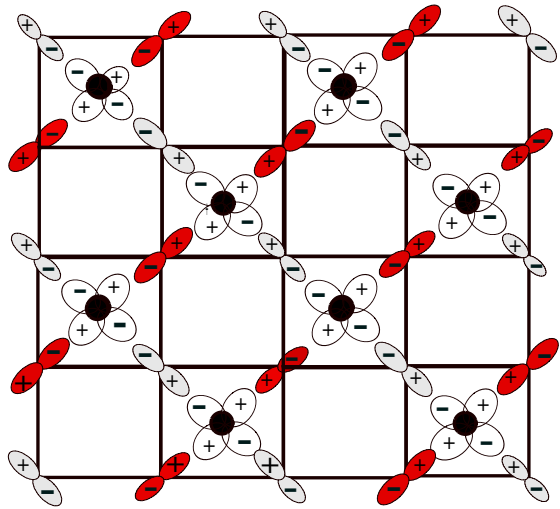}
	}
	\caption{The $CuO_2$ planar lattice with sub-lattices $A$ and $B$ represented in red and gray, respectively. This configuration, which corresponds to the base shown in Fig. \ref{f3},leads the magnetic interaction between localized and itinerant spins to an effective hole-attractive interaction for all nearest neighbor holes and is responsible for the superconductivity in cuprates. Notice that the base depicted in Fig \ref{f3} consists in a dimerization, if compared with the one in Fig. \ref{f1}.}
	\label{f4}
\end{figure}
\begin{figure}
	[h]
	\centerline{
		\includegraphics[scale=0.5]{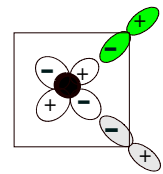}
	}
	\caption{The base of a possible crystal structure of the oxygen orbitals around the $Cu^{++}$ ions in the $CuO_2$ planar lattice.  This  configuration of oxygen orbitals leads the magnetic interaction between localized and itinerant spins to a hole-repulsive interaction for some of the nearest neighbor holes and to a hole-attractive interaction to the others. The black circles represent Cu ions, whereas the oxygen ions belonging to sub-lattices $A$ and $B$ are represented in green and gray, respectively. }
	\label{f1}
\end{figure}
\begin{figure}
	[h]
	\centerline{
		\includegraphics[scale=0.5]{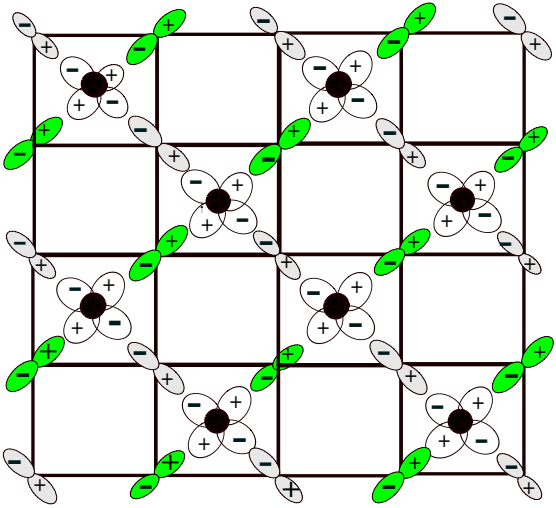}
	}
	\caption{The $CuO_2$ planar lattice with sub-lattices $A$ and $B$ represented in green and gray, respectively. This configuration, which corresponds to the base shown in Fig. \ref{f1},leads the magnetic interaction between localized and itinerant spins to produce hole-attractive and hole-repulsive interactions among the itinerant holes. }
	\label{f2}
\end{figure}
\begin{figure}
	[h]
	\centerline{
		\includegraphics[scale=0.35]{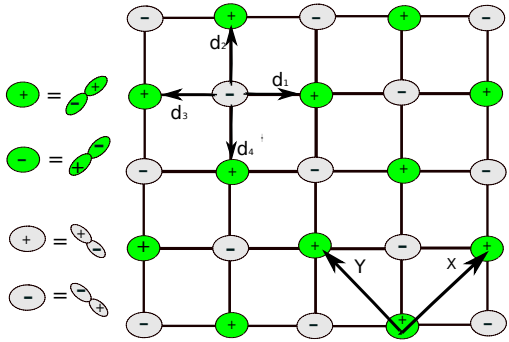}
	}
	\caption{The $A$ and $B$ sub-lattices, represented in green and gray, respectively. The sign of the p-orbitals is explicitly shown, indicating that each orbital of a given sub-lattice has four nearest neighbor orbitals belonging to the other one, indicated by $\textbf{d}_i, i=1,...,4$ with an opposite sign.}
	\label{figab}
\end{figure}

\bigskip

\begin{figure}
	[h]
	\centerline{
		\includegraphics[scale=0.35]{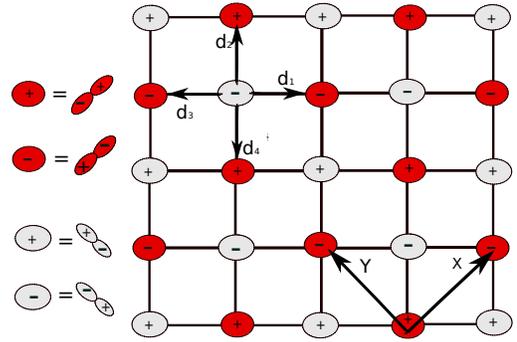}
	}
	\caption{The $A$ and $B$ sub-lattices, represented in red and gray, respectively. The sign of the p-orbitals is explicitly shown, indicating that each orbital of a given sub-lattice has four nearest neighbor orbitals belonging to the other one, indicated by $\textbf{d}_i, i=1,...,4$ such that the ones corresponding to $i=1,3$ have the same sign, while those corresponding to $i=2,4$  have a different sign.}
	\label{figab}
\end{figure}

According to Fig.\ref{f3}
and Fig.\ref{f4} (red configuration), we see that the oxygen orbitals can arrange themselves in such a way that $\eta_A\eta_C$ and $\eta_B\eta_C'$ have {\it opposite} signs for all nearest neighbor $(A,B)$ pairs.

This fact will be crucial for the magnetic interaction existing between the itinerant oxygen holes and the localized copper ions to produce a net {\it attractive} interaction between the holes belonging, respectively, to the $A$ and $B$ sub-lattices. At the same time this arrangement of the signs of the oxygen orbitals guarantees the super-exchange coupling that governs the magnetic interaction among the $Cu$-ion spins to be  {\it antiferromagnetic}. We can see clearly, therefore, that a close relation exists between the occurrence of a N\' eel state in the parent compound and the emergence of a superconducting state upon doping.
The interaction between the copper and oxygen spins resembles the one found in Kondo systems, in the sense that it couples the spins of itinerant holes with the spins of localized copper ions.
Notice, however, the profound difference that exists between the physics of the Kondo effect and that of the high-Tc cuprates: in the first case the itinerant electrons are essentially free, whereas in the present case 
the holes belong to the $p_x,p_y$ oxygen orbitals.

In the case of the cuprates, the holes, rather than being free, belong to a highly organized network of atomic p-orbitals of the oxygen ions. We shall see that there is a peculiar organization of the signs of such oxygen orbitals, that generates an effective attractive interaction between the holes belonging to the two $(A,B)$ oxygen sub-lattices thereby leading to a superconducting state. Below Tc, such arrangement, which leads to a superconductor state is, energetically, most favorable. \\
\bigskip
\vfill
\eject
{\bf 2.3) The Third Generation Effective Hamiltonian}\\

The above Hamiltonian does not describe the Coulomb repulsion between holes, however, given its relative magnitude (for LSCO $U_p =5.5 eV, J_{AF} =0.43 eV, J_K=1.17 eV$), it is natural to include Hubbard repulsive terms for the holes. This leads us to a third generation Hamiltonian, namely, the Spin-Fermion-Hubbard model \cite{emsc}, whose Hamiltonian possesses the four terms below: 

\begin{eqnarray}& &
\hspace{-5mm}H_0 =-t_{p} \sum_{\textbf{R},\textbf{d}_i}\sum_{\sigma}\psi_{A,\sigma}^\dagger(\textbf{R})\psi_{B,\sigma}(\textbf{R}+\textbf{d}_i) +hc
\nonumber \\
&\ &
\hspace{-5mm}H_U =
U_p \sum_{\textbf{R}} n^A_\uparrow n^A_\downarrow + U_p \sum_{\textbf{R}+\textbf{d}} n^B_\uparrow n^B_\downarrow 
\nonumber \\
& &
\hspace{-5mm}H_{AF} = J_{AF}\sum_{\langle IJ \rangle} \textbf{S}_I\cdot  \textbf{S}_J
\nonumber \\
& &
\hspace{-5mm}
H_K =J_K\sum_{I} \textbf{S}_I\cdot\left[\sum_{\textbf{R}\in I} \eta_A \eta_C\  \mathcal{S}_A +
\sum_{\textbf{R}+\textbf{d}\in I} \eta_B \eta_C' \    \mathcal{S}_B   \right ]
\nonumber \\
& &
\label{0ax}
\end{eqnarray}\\
\bigskip

{\bf 2.4) The Fourth Generation Effective Hamiltonian}\\

In order to derive the effective Hamiltonian describing the interaction among the doped holes, we are going to perform two operations on the previous Hamiltonian: 1) we are going to trace over the ferromagnetic fluctuations component of localized spins $\textbf{S}_I$ in the last two terms; 2) in the first two terms, we are going to do a perturbation expansion on the hopping term with respect to the unperturbed Hubbard term. These two operations will lead to our effective Hamiltonian for the cuprates, which is a fourth generation Hamiltonian.

We start by tracing out the ferromagnetic fluctuations of the localized spins. This will produce an effective interaction Hamiltonian, $H_{1}[\psi]$, for the itinerant holes.

We have
\begin{eqnarray}
Z = \mathrm{Tr}_\psi\mathrm{Tr}_{\textbf{S}_I}e^{-\beta\left[ H_0[\psi] + H_U[\psi]+ H[\textbf{S}_I,\psi]\right]}
\label{976b}
\end{eqnarray}

We can express the trace over such localized spins $\textbf{S}_I$ as a functional integral over the classical vector field $\textbf{N}$ of unit length (see, for instance \cite{M1}) which is introduced through an appropriate base of coherent spin states, namely:
\begin{eqnarray}
 \langle \textbf{N} |\textbf{S}_I| \textbf{N} \rangle =s \textbf{N}_I
\label{976b}
\end{eqnarray}
($s=1/2$ is the spin quantum number)
from which, we have \cite{ecm2}
\begin{eqnarray}
\mathrm{Tr}_{\textbf{S}_I}
e^{-\beta H[\textbf{S}_I,\psi]}
=
\int D \textbf{N}
\delta(|\textbf{N}|^2-1)e^{-\beta H[s\textbf{N}_I,\psi]}
\label{976bc}
\end{eqnarray}

We now consider separately the antiferromagnetic and ferromagnetic components of $\textbf{N}$, denoted, respectively, by
 $\textbf{n}$ and $\textbf{L}$. We then write, expanding up to the second order in the lattice parameter $a$ 
 \begin{eqnarray}
\textbf{N}_I= (-1)^I \textbf{n}_I+a^2\textbf{L}_I+O(a^4)
\label{988}
\end{eqnarray}
such that $|\textbf{N}|^2=|\textbf{n}|^2=1$ and 
$\textbf{L}\cdot\textbf{n}=0$. 

We can re-write the trace over $\textbf{S}_I$ as a double functional integral on  $\textbf{n}$ and $\textbf{L}$ \cite{M1}:
\begin{eqnarray}
&\ &\hspace{-5mm}\mathrm{Tr}_{\textbf{S}_I}= \int D\textbf{n}D\textbf{L}\delta(|\textbf{n}|^2-1)
\label{nL}
\end{eqnarray}

\begin{eqnarray}
&\ &\hspace{-5mm}\mathrm{Tr}_{\textbf{S}_I}e^{-\beta H[\textbf{S}_I,\psi]}= \int D\textbf{n}D\textbf{L}\delta(|\textbf{n}|^2-1)
\nonumber \\
&\ &\hspace{-5mm} \exp\left\{-\frac{1}{2}\int d^2r\int_0^\beta d\tau\left[ J_{AF}s^2 \nabla_i \textbf{n}\cdot \nabla_i \textbf{n} \right. \right .
\nonumber \\
&\ &\hspace{-5mm}\left .+ 4J_{AF}s^2a^2  |\textbf{L}|^2\right .
\nonumber \\
&\ & \left .
+ \textbf{L}\cdot\left[ J_K \left[\eta_A \eta_C\  \mathcal{S}_A +\eta_B \eta_C' \    \mathcal{S}_B  \right ]
     -is\textbf{n}\times\frac{\partial \textbf{n} }{\partial \tau} \right ]  \right\}
\label{99110}
\end{eqnarray}

We are going to integrate out the ferromagnetic fluctuations by performing the quadratic functional integral on $\textbf{L}$. This will produce three terms: the 2nd term squared, which provides a kinetic term for $\textbf{n}$ \cite{M1}, the 1st term squared, which produces an effective interaction among the itinerant doped holes and the crossed term, which vanishes \cite{em,ecm2}.
\begin{widetext}

\begin{eqnarray}
& &
\mathrm{Tr}_{\textbf{S}_I}e^{-\beta H[\textbf{S}_I,\psi]}
=\int D\textbf{n}\delta(|\textbf{n}|^2-1) \times
\nonumber \\
& &\exp\left\{-\int d^2r \int_0^\beta d\tau\frac{\rho_s}{2}\left[  \nabla_i \textbf{n}\cdot \nabla_i \textbf{n} +\frac{1}{c^2} \partial_\tau \textbf{n}\cdot \partial_\tau \textbf{n}
\right]\right .
\nonumber \\
& &+\frac{J^2_K}{8J_{AF}a^2}\left[\sum_{\textbf{R}\in I} \eta_A \eta_C\  \mathcal{S}_A +
\sum_{\textbf{R}+\textbf{d}\in I} \eta_B \eta_C' \    \mathcal{S}_B   \right ]\cdot \left[\sum_{\textbf{R}\in I} \eta_A \eta_C\  \mathcal{S}_A +
\sum_{\textbf{R}+\textbf{d}\in I} \eta_B \eta_C'\    \mathcal{S}_B   \right ]   
\left .\right \}
\label{99113}
\end{eqnarray}
where $\rho_s = \frac{J_{AF}}{4}$ is the spin stiffness and $c=J_{AF}a$ is the spin-waves velocity.

Using the fact that the continuum limit involves the replacement
$a^2 \sum_k \leftrightarrow \int d^2r $, we conclude that
\begin{eqnarray}
 {\rm Tr}_{\textbf{S}_I} e^{- \beta\Big[ H_{AF}[\textbf{S}_I]+H_{K}[\textbf{S}_I,\psi]\Big]} =Z_{NL\sigma M} e^{- \beta  H_{1}[\psi]}=
& Z_{NL\sigma M}\exp\left\{-\int_0^\beta d\tau\sum_{\textbf{R},\textbf{R}+\textbf{d}} \left[ \frac{J^2_K}{8J_{AF}} \left[\eta_A\eta_C\mathcal{S}_A +\eta_B\eta'_C\mathcal {S}_B\right]^2\right ]\right\},
\label{976b}
\end{eqnarray}
where $Z_{NL\sigma M}$ is the partition function of the Nonlinear Sigma Model ( see, for instance \cite{ecm2}).

From the last term in (\ref{976b}) we see that, indeed,
\begin{eqnarray}
H_{1}[\psi]=\frac{J^2_K}{8J_{AF}}\sum_{\textbf{R},\textbf{R}+\textbf{d}}  [\eta_A\eta_C\mathcal{S}_A +\eta_B\eta_C'\mathcal{S}_B]^2.
\label{h1}
\end{eqnarray}

Inserting the expressions for $\mathcal{S}_A$ and $\mathcal{S}_B$ in (\ref{h1}), we obtain, up to a constant,
\begin{eqnarray}
&\ &
H_{1}[\psi]=
\nonumber \\
&\ &\frac{J^2_K}{8J_{AF}}\eta_A\eta_B\eta_C\eta_C'
\sum_{\textbf{R},\textbf{R}+\textbf{d}_i} 
\Big [\psi_{A\uparrow}^\dagger(\textbf{R})\psi_{B\downarrow}^\dagger(\textbf{R}+\textbf{d}_i)
 + \psi_{A\downarrow}^\dagger(\textbf{R})
\psi^\dagger_{B\uparrow}(\textbf{R}+\textbf{d}_i)\Big]
\Big [\psi_{B\downarrow}(\textbf{R}+\textbf{d}_i)\psi_{A\uparrow}(\textbf{R})
 +  \psi_{B\uparrow}(\textbf{R}+\textbf{d}_i) \psi_{A\downarrow}(\textbf{R})\Big]
\nonumber \\
& &\hspace{-3mm}
\label{h1a}
\end{eqnarray}
\end{widetext}
From Fig. \ref{f1} we see that
\begin{equation}
\eta_A\eta_C\eta_B\eta_C'=-1.
\label{1a}
\end{equation}
For this reason,
the above interaction is always attractive between nearest neighbor holes, which belong to different sub-lattices.

The complete effective Hamiltonian for the cuprates is obtained by carrying on a second order perturbation theory on the $H_0 + H_U$ terms \cite{M1}. Including this result, we obtain the following, fourth generation effective Hamiltonian:
\begin{widetext}

\begin{eqnarray}
&\ &\hspace{-3mm}H_{eff}[\psi]=-t \sum_{\textbf{R},\textbf{d}_i} \psi_{A\sigma}^\dagger(\textbf{R})\psi_{B\sigma}(\textbf{R}+\textbf{d}_i)+hc
\nonumber \\
&\ &\hspace{-3mm}-g_S\sum_{\textbf{R},\textbf{d}_i} \Big [\psi_{A\uparrow}^\dagger(\textbf{R})\psi_{B\downarrow}^\dagger(\textbf{R}+\textbf{d}_i)
 + 
\psi^\dagger_{B\uparrow}(\textbf{R}+\textbf{d}_i)\psi_{A\downarrow}^\dagger(\textbf{R})\Big]
\Big [\psi_{B\downarrow}(\textbf{R}+\textbf{d}_i)\psi_{A\uparrow}(\textbf{R})
 +  \psi_{A\downarrow}(\textbf{R}) \psi_{B\uparrow}(\textbf{R}+\textbf{d}_i)\Big]
\nonumber \\
&\ &\hspace{1mm}-g_P
\sum_{\textbf{R},\textbf{d}_i} \Big [\psi_{A\uparrow}^\dagger(\textbf{R})\psi_{B\uparrow}(\textbf{R}+\textbf{d}_i)
 +\psi_{A\downarrow}^\dagger(\textbf{R}) \psi_{B\downarrow}(\textbf{R}+\textbf{d}_i)
\Big]
\hspace{1mm}
\Big [\psi_{B\uparrow}^\dagger(\textbf{R}+\textbf{d}_i)\psi_{A\uparrow}(\textbf{R})
+\psi_{B\downarrow}^\dagger(\textbf{R}+\textbf{d}_i) \psi_{A\downarrow}(\textbf{R})\Big],
\label{0}
\end{eqnarray}
 where $g_S$, is the hole-attractive interaction coupling parameter and $g_P$, the hole-repulsive one, given, respectively, by
\begin{eqnarray}
 g_S = \frac{ J^2_K }{8 J_{AF}}   \ \ \ \; \ \ \  g_P = \frac{2 t_p^2}{U_p}
\label{01}
\end{eqnarray}

\vfill
\eject
\end{widetext}
{\bf 3) The Effective Hamiltonian}\\
 \bigskip

{\bf 3.1) Hubbard-Stratonovitch Fields}\\
\bigskip

We can write our effective Hamiltonian in terms of the Hubbard-Stratonovitch fields 
$\Phi$ and $\chi$, as \cite{M1}
\begin{widetext}
\begin{eqnarray}
&\ & \hspace{-3mm}H_{eff}=-t_{p} \sum_{\textbf{R},\textbf{d}_i}\sum_{\sigma}\psi_{A,\sigma}^\dagger(\textbf{R})\psi_{B,\sigma}(\textbf{R}+\textbf{d}_i) +hc
\nonumber \\
&\ &\hspace{-3mm}
+ \sum_{\textbf{R},\textbf{d}_i}\Phi(\textbf{R},\textbf{d}_i) 
\Big[\psi_{A\uparrow}^\dagger(\textbf{R})\psi_{B\downarrow}^\dagger(\textbf{R}+\textbf{d}_i)
+\psi^\dagger_{B\uparrow}(\textbf{R}+\textbf{d}_i) \psi_{A\downarrow}^\dagger(\textbf{R})
\Big]+ hc
\nonumber\\
&\ &\hspace{-3mm}
+\sum_{\textbf{R},\textbf{d}_i}\chi(\textbf{R},\textbf{d}_i) \Big [\psi_{A\uparrow}^\dagger(\textbf{R})\psi_{B\uparrow}(\textbf{R}+\textbf{d}_i)
 + \psi_{A\downarrow}^\dagger(\textbf{R})\psi_{B\downarrow}(\textbf{R}+\textbf{d}_i)\Big]+ hc
 \nonumber \\
&\ &\hspace{-3mm}+\frac{1}{g_S}\sum_{\textbf{R},\textbf{d}_i}\Phi^\dagger(\textbf{R},\textbf{d}_i) \Phi(\textbf{R},\textbf{d}_i)
+\frac{1}{g_P}\sum_{\textbf{R},\textbf{d}_i\in \textbf{R}}\chi^\dagger(\textbf{R},\textbf{d}_i)\chi(\textbf{R},\textbf{d}_i) ,
\label{1a}
\end{eqnarray}

In order to describe the doping process, we add to the above Hamiltonian the chemical potential term
\begin{eqnarray}
-\mu\left[\sum_{\sigma}\left ( \psi^{\dagger}_{A,\sigma}\psi_{A,\sigma} + \psi^{\dagger}_{B,\sigma}\psi_{B,\sigma}\right)-d(x)\right]
\end{eqnarray}
where $d(x)$ is a function of the stoichiometric doping parameter $x$, to be determined self-consistently.
\end{widetext}
From this we derive the field equations

\begin{eqnarray}
& &\Phi^\dagger(\textbf{R},\textbf{d}_i)= 
\\ \nonumber
& &g_S\Big[\psi_{A\uparrow}^\dagger(\textbf{R})\psi_{B\downarrow}^\dagger(\textbf{R}+\textbf{d}_i)
+\psi^\dagger_{B\uparrow}(\textbf{R}+\textbf{d}_i) \psi_{A\downarrow}^\dagger(\textbf{R})
\Big]
\label{1b}
\end{eqnarray}
and
\begin{eqnarray}
& &\chi^\dagger (\textbf{R},\textbf{d}_i) =  
\\ \nonumber
& &g_P\Big [\psi_{A\uparrow}^\dagger(\textbf{R})\psi_{B\uparrow}(\textbf{R}+\textbf{d}_i)
 + \psi_{A\downarrow}^\dagger(\textbf{R})\psi_{B\downarrow}(\textbf{R}+\textbf{d}_i)\Big]
\label{1c}
\end{eqnarray}

Observe that
$\Phi^\dagger$ creates two holes with opposite spins, each one in the neighboring sites $A,B$, located, respectively, at $(\textbf{R},\textbf{R}+\textbf{d}_i)$, where
\begin{eqnarray}
& &\textbf{d}_1=\frac{1}{2}[\textbf{X}-\textbf{Y}]\ ;\ \textbf{d}_2=\frac{1}{2}[\textbf{X}+\textbf{Y}]\nonumber \\
& &\textbf{d}_3=\frac{1}{2}[-\textbf{X}+\textbf{Y}]\ ;\ \textbf{d}_4=\frac{1}{2}[-\textbf{X}-\textbf{Y}] 
\label{11xa},
\end{eqnarray} 
where $\textbf{X}=a \hat{x}$ and $\textbf{Y}=a \hat{y}$ are primitive vectors of the oxygen lattices (see Fig. \ref{figab}). It is, therefore, a Cooper pair creation operator.  $\chi^\dagger$, conversely, creates an electron and a hole with parallel spins, also in such neighboring sites $A,B$. It is, therefore, an exciton creation operator. \\
\bigskip
\vfill
\eject
{\bf 3.2) The SC and PG Order Parameters}\\
\bigskip

Let us examine the ground-state expectation value of these operators, namely
\begin{eqnarray}
& &\Delta(\textbf{d}_i)= 
\\ \nonumber
& &g_S\Big\langle\psi_{A\uparrow}^\dagger(\textbf{R})\psi_{B\downarrow}^\dagger(\textbf{R}+\textbf{d}_i)
+\psi^\dagger_{B\uparrow}(\textbf{R}+\textbf{d}_i) \psi_{A\downarrow}^\dagger(\textbf{R})
\Big\rangle
\label{1b}
\end{eqnarray}
and
\begin{eqnarray}
& &  M(\textbf{d}_i) =  
\\ \nonumber
& &g_P\Big \langle\psi_{A\uparrow}^\dagger(\textbf{R})\psi_{B\uparrow}(\textbf{R}+\textbf{d}_i)
 + \psi_{A\downarrow}^\dagger(\textbf{R})\psi_{B\downarrow}(\textbf{R}+\textbf{d}_i)\Big\rangle
\label{1c}
\end{eqnarray}
Notice that, because of the invariance under Bravais lattice translations, it follows that $\Delta(\textbf{d}_i)$ and $M(\textbf{d}_i)$ do not depend on the Bravais lattice sites' position $\textbf{R}$, rather, they depend only on $\textbf{d}_i$.

According to Fig. \ref{figab}, we see that
\begin{eqnarray}
\Delta(\textbf{d}_{1,3})=- \Delta(\textbf{d}_{2,4}) 
\label{d}
\end{eqnarray}
and
\begin{eqnarray}
M(\textbf{d}_{1,3})= - M(\textbf{d}_{2,4}) . 
\label{m}
\end{eqnarray}
Now, considering that
\begin{eqnarray}
\Delta(\textbf{k})=\sum_{\textbf{d}_{i}=1,...,4} \Delta(\textbf{d}_i)\exp\{i \textbf{k}\cdot \textbf{d}_{i} \}
\label{1bb}
\end{eqnarray}
and that
\begin{eqnarray}
M(\textbf{k})=\sum_{\textbf{d}_{i}=1,...,4}M(\textbf{d}_i) \exp\{i \textbf{k}\cdot \textbf{d}_{i} \}, 
\label{1cc}
\end{eqnarray}
we find, using the fact that $|\textbf{d}_i|=a'= a/\sqrt{2}$, 
\begin{eqnarray}
\Delta(\textbf{k}) = \Delta\left[\cos k_+a' - \cos k_- a' \right]
\label{1bb}
\end{eqnarray}
and also that
\begin{eqnarray}
M(\textbf{k})= M \left[\cos k_+a' - \cos k_- a' \right]
\label{1cc}
\end{eqnarray}
where $k_\pm=\frac{k_x \pm k_y}{\sqrt{2}}$.

As it was shown in \cite{M1}, $M$ is pure imaginary.

We see that the SC and PG order parameters both have a d-wave symmetry, namely, change the sign under a $90^\circ$ rotation, and have nodal lines along the $\pm \hat{x}$ and $\pm \hat{y}$ directions.\\

\bigskip
{\bf 4) The Spectral Weight and the PG Order Parameter}\\
\bigskip

{\bf 4.1) General Expression}\\
\bigskip
Consider a system with eigen-energies given by $E_n=\hbar \omega \left(\textbf{k}_n\right)$.
The spectral weight for such a system is defined
as \cite{ecm2}
\begin{eqnarray}
& &N(\omega)= \sum_{\textbf{k}_n}\delta\left(\omega-\omega \left (\textbf{k}_n\right)\right )
\nonumber \\
& &N(\omega)=-\frac{1}{\pi} \mathrm{Im}
\sum_{\textbf{k}_n}\frac{1}{\omega -\omega \left (\textbf{k}_n\right) +i \epsilon }
\label{1}
\end{eqnarray}

This can be written as 
\begin{eqnarray}
N(\omega)= \int_{-\infty}^{\infty} \frac{d\lambda}{2\pi}e^{-i\lambda\omega}\sum_{\textbf{k}_n} e^{i\lambda\omega \left (\textbf{k}_n\right)}
\label{2}
\end{eqnarray}

or, equivalently,
\begin{eqnarray}
N(\omega)= \int_{-\infty}^{\infty} \frac{d \lambda}{2\pi}e^{-i\lambda\omega}f(\lambda)
\label{3}
\end{eqnarray}
where
\begin{eqnarray}
f(\lambda)=\sum_{\textbf{k}_n} e^{i\lambda\omega \left (\textbf{k}_n\right)}.
\label{4}
\end{eqnarray}
In the case of a continuum spectrum, this is given by
\begin{eqnarray}
f(\lambda)=\int \frac{d^2 k}{(2\pi)^2} e^{i\lambda\omega \left (\textbf{k}\right)}.
\label{5}
\end{eqnarray}\\

\bigskip

\vfill
\eject

{\bf 4.2) The Spectral Weight in the Strange Metal Phase}\\

\bigskip
In the SM phase, we have $M=0$, and the dispersion relation is given by
\begin{eqnarray}
\omega(\textbf{k})=  V\left[\cos \frac{k_+a}{\sqrt{2}} + \cos \frac{k_-a}{\sqrt{2}}\right ]=2V\cos \frac{k_x a}{\sqrt{2}}\cos \frac{k_y a}{\sqrt{2}}
\\ \nonumber
\label{a1}
\end{eqnarray}

Expanding around the Brillouin zone center, we have
\begin{eqnarray}
\omega(\textbf{k})= 2V\left[1-\frac{a^2}{4}\left(k_x^2 +k_y^2 \right )\right ]\label{a2}
\end{eqnarray}

Inserting this in (\ref{5}), and using the result
\begin{eqnarray}
 \int_{-\infty}^{\infty} \frac{d \lambda}{2\pi}e^{-i\lambda\left(\omega-\omega_0\right )}= i\pi-2\pi i\theta(\omega-\omega_0),
 \label{a3}
\end{eqnarray}
where $\theta(x)$ is the Heaviside function, we obtain the spectral weight in the SM phase:
\begin{eqnarray}
 N(\omega)= \left( \frac{4}{a^2}  \right )\left[ \theta(\omega)-\theta(\omega-\omega_F)  \right ]
 \label{a4}
\end{eqnarray}\\

\bigskip

\vfill\eject
{\bf 4.3) The Spectral Weight in the Pseudogap Phase}\\

\begin{figure}
	[!h]
	\centerline{
		\includegraphics[scale=0.25]{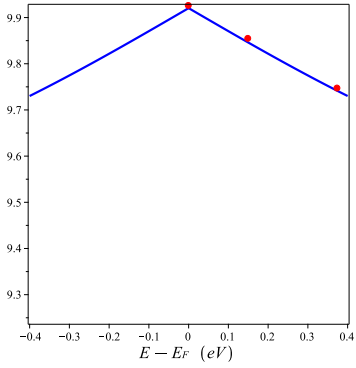}
	}
	\caption{The Spectral Weight for Bi2212 at $T=T^*=190K$, namely, at the boundaray between the SM and PG phases (arbitrary units). Experimental data from \cite{sw}. Solid line is a plot of (\ref{a14}), in the limit where $M\rightarrow 0$.}
	\label{fig1}
\end{figure}

\begin{figure}
	[!h]
	\centerline{
		\includegraphics[scale=0.25]{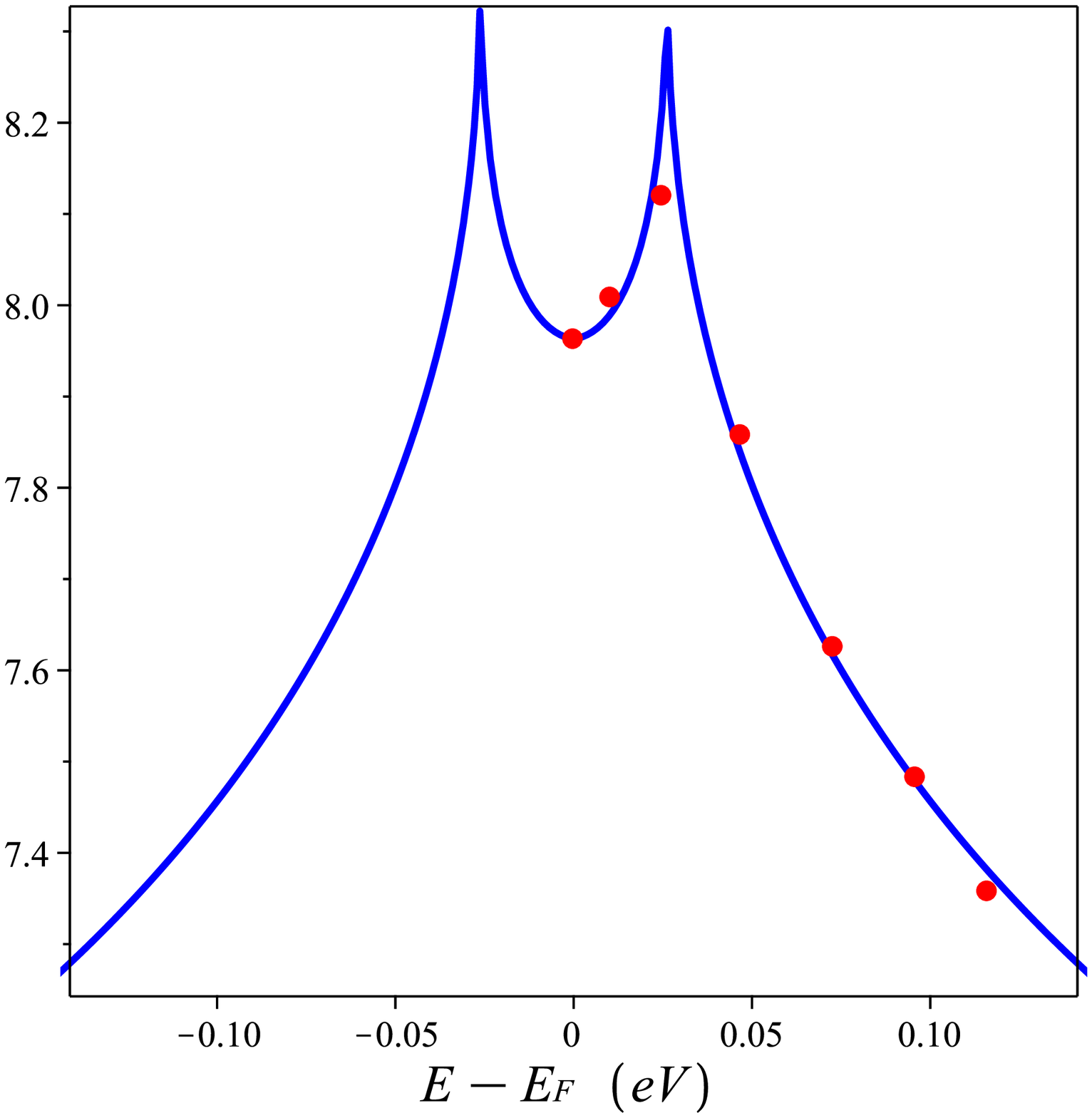}
	}
	\caption{The Spectral Weight in the PG Phase for Bi2212 at 144K (arbitrary units). Experimental data from \cite{sw}. Solid line is a plot of (\ref{a14}).}
	\label{fig2}
\end{figure}

\begin{figure}
	[!h]
	\centerline{
		\includegraphics[scale=0.25]{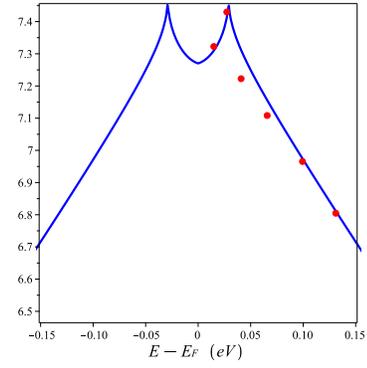}
	}
	\caption{The Spectral Weight in the PG Phase for Bi2212 at 123K (arbitrary units). Experimental data from \cite{sw}. Solid line is a plot of (\ref{a14}).}
	\label{fig3}
\end{figure}



\bigskip
\begin{widetext}
In the PG phase,we have $M\neq 0$ and considering just the two oxygen sub-lattices, we have
\begin{eqnarray}
\omega(\textbf{k})= \sqrt{  V^2\left[\cos k_xa + \cos k_ya \right ]^2 +  M^2\left[\cos k_xa - \cos k_ya \right ]^2  }.
\\ \nonumber
\label{a5}
\end{eqnarray}

Now, making an expansion analogous to that we did before, but now expanding around the center of the elliptical Fermi surface pockets, namely, $\mathbf{K}=(\frac{\pi}{2a},\frac{\pi}{2a})$\cite{M1} we obtain
\begin{eqnarray}
\omega(\textbf{k})\simeq \sqrt{2}\left[\frac{V^2 +M^2}{\sqrt{V^2 +M^2}}\right]\frac{k^2a^2}{4}+
\left[\frac{V^2-M^2}{\sqrt{V^2+M^2}}\right] k^2a^2 \sin(2\theta).
\label{a7}
\end{eqnarray}
where $y\equiv k^2=k_x^2 +k_y^2$, $k_x=k\cos\theta$ and $k_y = k\sin\theta$.

Considering that the Bessel function $J_0(x)$ is given by
\begin{eqnarray}J_0(x)=
\frac{2}{\pi}\int_{0}^{\pi}d\theta\cos\Big(x\sin\theta\Big),
\label{a9}
\end{eqnarray}
and using (\ref{5}), we arrive, after performing the $\theta$-angular integration, at
\begin{eqnarray}
N(\omega) =\frac{1}{4\pi a^2} \int_0^{k^2_F}d y \int_{-\infty}^{\infty} \frac{d \lambda}{2\pi}
\exp\left[-i\lambda \omega + i\lambda \Big[\left[\frac{V^2 +M^2}{\sqrt{V^2 +M^2}}\right]\frac{ya^2}{4}\Big]\right ]
J_0\left ( \lambda \left[\frac{V^2 -M^2}{\sqrt{V^2 +M^2}}\right]\frac{ya^2}{4} \right ).
\label{a10}
\end{eqnarray}
\begin{figure}
	[!h]
	\centerline{
		\includegraphics[scale=0.25]{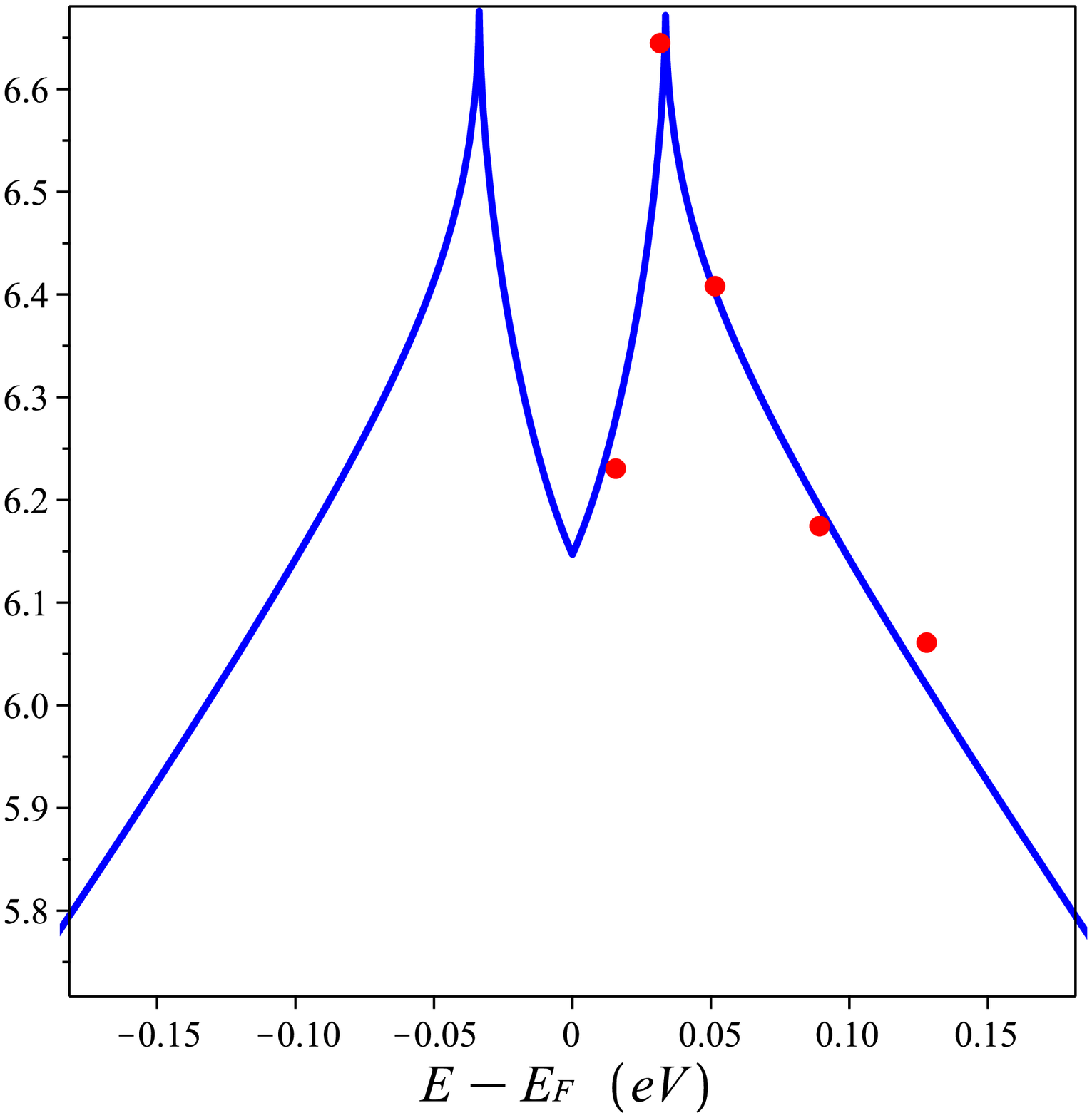}
	}
	\caption{The Spectral Weight in the PG Phase for Bi2212 at 100K (arbitrary units). Experimental data from \cite{sw}. Solid line is a plot of (\ref{a14}).}
	\label{figx4}
\end{figure}
\begin{figure}
	[h]
	\centerline{
		\includegraphics[scale=0.25]{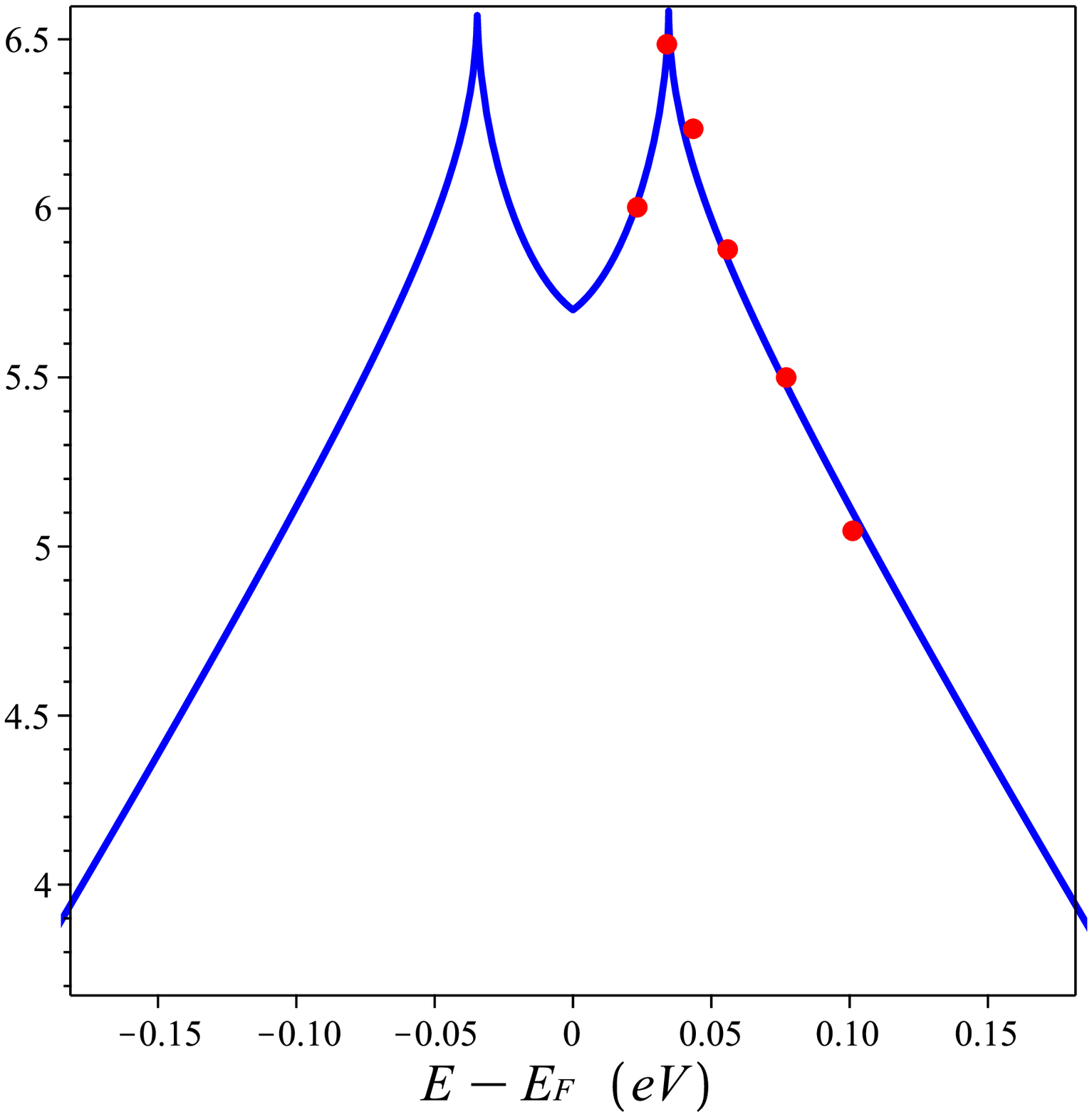}
	}
	\caption{The Spectral Weight in the PG Phase for Bi2212 at 92K (arbitrary units). Experimental data from \cite{sw}. Solid line is a plot of (\ref{a14}).}
	\label{fig5}
\end{figure}
Now, integrating over $\lambda$ \cite{gradshteyn}, 
we obtain
\begin{eqnarray}
&\ &= \frac{1}{4\pi a^2} \int_0^{k^2_F}dy
\frac{1}{\sqrt{\Big [ \frac{2M^2}{\sqrt{V^2+M^2}} y -\omega\Big ]\Big [\frac{2M^2}{\sqrt{V^2+M^2}}y +\omega\Big]}}
\label{a11}
\end{eqnarray}

\end{widetext}

Finally, integrating on $y$, we arrive at
\begin{eqnarray}
N(\omega ) \propto
\arctan\Big \{\frac{1}{\bigskip{|A^2-\omega^2|}}\Big\}-\arctan\Big \{\frac{1}{|\omega|} \Big \}
\label{a14}
\end{eqnarray} 
where $A=\frac{2M^2k^2_F}{\sqrt{V^2 +M^2}}$. This function has sharp peaks at $\pm A$, as we can see in Figures \ref{fig2},...,\ref{fig5}. The peaks occur at frequencies which are proportional to the pseudogap. The opening of a region of lesser spectral weight in between the two peaks has been observed in the whole PG phase by means of tunneling experiments and is a convincing experimental proof of the correctness of our prediction about the existence of a PG order parameter in the whole PG phase of the cuprates: the PG order parameter is proportional to the peaks separation.\\

{\bf 5) The SC and PG Transition Temperatures: $T_c$ and $T^*$}\\
\bigskip
{\bf 5.1) The SC Transition Temperatures: $T_c$ }\\
\bigskip
Starting from the effective Hamiltonian (\ref{1a}) and integrating over the fermion fields, we obtain an effective grand-partition function, $\Omega(\Delta, M,\mu)$, which depends on the SC and PG order parameters, as well as on the chemical potential \cite{M1}.
Imposing the stationary condition on $\Delta$ and $M$, we obtain, respectively, the equations determining the transition temperatures $T_c$ and $T^*$, which are given by

\begin{equation}
		\begin{cases}
			T_{c}(x) =\frac{\ln2 \,\ T_{max}}{\ln2 + \frac{\mu_0(x)}{2T_{c}(x)} - \frac{1}{2}\left(1-e^{-\frac{\mu_0(x)}{T_{c}(x)}}\right)},\hspace{0.5cm}x < x_{0}\\
			T_c(x) =\frac{\ln2 \ \ T_{max}}{\ln\Big [1+ \exp\left[- \frac{\mu_0(x)}{T_c(x)} \right]  \Big ]},\hspace{1.6cm} x > x_{0}\\
			T_{c}(x) =\frac{\ln2 \,\ T_{max}}{\ln2 + \frac{|\mu_0(x)|}{2T_{c}(x)} - \frac{1}{2}\left(1-e^{-\frac{|\mu_0(x)|}{T_{c}(x)}}\right)},\hspace{0.3cm} \text{LSCO},\forall x
			\label{eqtc}
		\end{cases}
	\end{equation}
	
	The chemical potential at the curve $T(x)$, is expressed as
\begin{equation}
\mu_0(x)=2\gamma(x_0-x),
			\label{mu}
	\end{equation}
 and $\gamma$ is a parameter which must be determined for each compound. $T_{max}$ is the transition temperature at optimal doping, $x_0$, which is given by 
	
	\begin{equation}
	    T_{max} =\frac{\Lambda \eta (Ng_S)}{2\ln 2}
	\end{equation}
where $\eta\left(N g_S\right)=1-\frac{g_c}{N g_S}$,

where $\Lambda=0.018$ eV is a characteristic  energy scale. 

We have the SC dome ending at the points $x^\pm_{SC}$ at the left and right, where
\begin{equation}
	    x^+_{SC}=x_0 +\frac{\Lambda \eta}{4\gamma}\ \ \ ;\ \ \ x^-_{SC}=x_0 -\frac{\Lambda \eta}{2\gamma}
	\end{equation}

Observe that for LSCO we use a symmetrized version of the equations to comply with the experimental observation that the SC dome is symmetrical for this compound.

{\bf 5.2) The PG Transition Temperatures: $T^*$ }\\
\bigskip
The PG transition temperature satisfies the following equation
	\begin{equation}
		T^*(x) =\frac{\frac{\Lambda\tilde\eta(g_PN)}{2}}{\ln\Big [1+ \exp\left[- \frac{\tilde\mu(x)}{T^*(x)} \right]  \Big ]},
		\label{eq_Ts}
	\end{equation}

	In the above equation 
	\begin{equation}
\tilde\mu(x)=2\tilde\gamma (\tilde x_0 - x),
			\label{mutil}
	\end{equation}
	is the chemical potential at the curve $T^*(x)$.
		 $\tilde x_0$ is a parameter determining the point where $T^*\rightarrow 0 $ and $\tilde\gamma$ must be determined for each compound \cite{M1}. 

 Calling $x^+_{PG}$ the point where the curve $T^*(x)$ reaches zero, namely, $T^*(x^+_{PG})=0$, we have
 \begin{equation}
x^+_{PG} = \tilde x_0 +\frac{\Lambda \tilde \eta}{4 \tilde \gamma}
			\label{xpg}
	\end{equation}

Inserting the values of the quantities in the above equation, we can see that $\frac{\Lambda \tilde \eta}{4 \tilde \gamma}\ll \tilde x_0$, hence $x^+_{PG} \simeq \tilde x_0 $.

\bigskip

{\bf 5.3) The Relation $\gamma x_0 \eta^{1/N} = \tilde\gamma \tilde x_0 \tilde\eta^{1/N}$}\\
\bigskip

By considering the stationary condition on the chemical potential, namely,
\begin{equation}
    \frac{\partial \Omega(\Delta, M,\mu)}{\partial \mu}=0, 
\label{pm}
\end{equation}
we obtain \cite{M1}, for the case of systems with one $CuO_2$-plane per unit cell (N=1), 
\begin{equation}
    d(x) =2\mu_0 (x) \frac{\eta(N=1)}{g_c}
    \label{m1}
\end{equation}
The optimal doping occurs at $x_0$ such that $\mu(x=x_0) =0$, whence we write the chemical potential in the form 
expressed in (\ref{mu}).

This implies
\begin{equation}
  d(0) =\frac{4\gamma x_0}{g_c} \eta(N=1)  
       \label{m2}
\end{equation}
Repeating the same sequence of arguments for the system right over the $T^*(x)$  curve, we obtain the relation
\begin{equation}
    d(0) \frac{g_c}{4}=\gamma x_0 \eta(N=1)=\tilde\gamma \tilde x_0 \tilde\eta(N=1),
    \label{m3}
\end{equation}
which we had obtained previously \cite{M1}.

In the case of $N$ multiple $CuO_2$ planes, we will have $N$ functions $d_i(x): i=1,...,N$ and one finds, instead, for the $i^{th}$ plane, 
\begin{equation}
 [d_i(0)]^N =\left[\frac{4\gamma x_0}{g_c}\right]^N \eta^N_i\ \ \ ;\ \ \  \eta_i=\eta(1)^{1/N} \simeq \eta(N)
     \label{m4}
\end{equation}
which implies
\begin{equation}
    d_i(0) \frac{g_c}{4}=\gamma x_0 \eta(N=1)^{1/N}
    \label{m33}
\end{equation}
and, consequently, by the same reasoning above, we have
\begin{equation}
    \left[(\gamma x_0) \eta(1)^{1/N}\right]=\left[(\tilde\gamma \tilde x_0) \tilde\eta(1)^{1/N}\right].
    \label{m5}
\end{equation}

\begin{figure}
	[!h]
	\centerline{
		\includegraphics[scale=0.3]{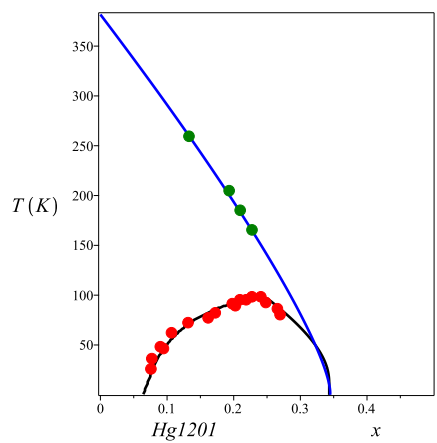}}
		\caption{Hg1201. Experimental data from \cite{01}}
	\label{f03}
\end{figure}

\begin{widetext}

In Tables 1 and 2, we list the values of the relevant parameters for several High-Tc cuprates. Observe, in particular, the value of the combinations appearing in (\ref{m5}).
In Figs. \ref{f03} to \ref{f09}, we display our theoretical SC
and PG transition lines for LSCO and for the members of the Bi and Hg families, along with the corresponding experimental data \cite{001,002,003}.
Notice that for single-layered 
materials, the phase-diagram graphs coincide with the ones in \cite{M1}, while for multi-layered ones ($N>1$), they differ, in compliance with the new relation (\ref{m5}).
\end{widetext}

\begin{figure}
	[h!]
	\centerline{
		\includegraphics[scale=0.3]{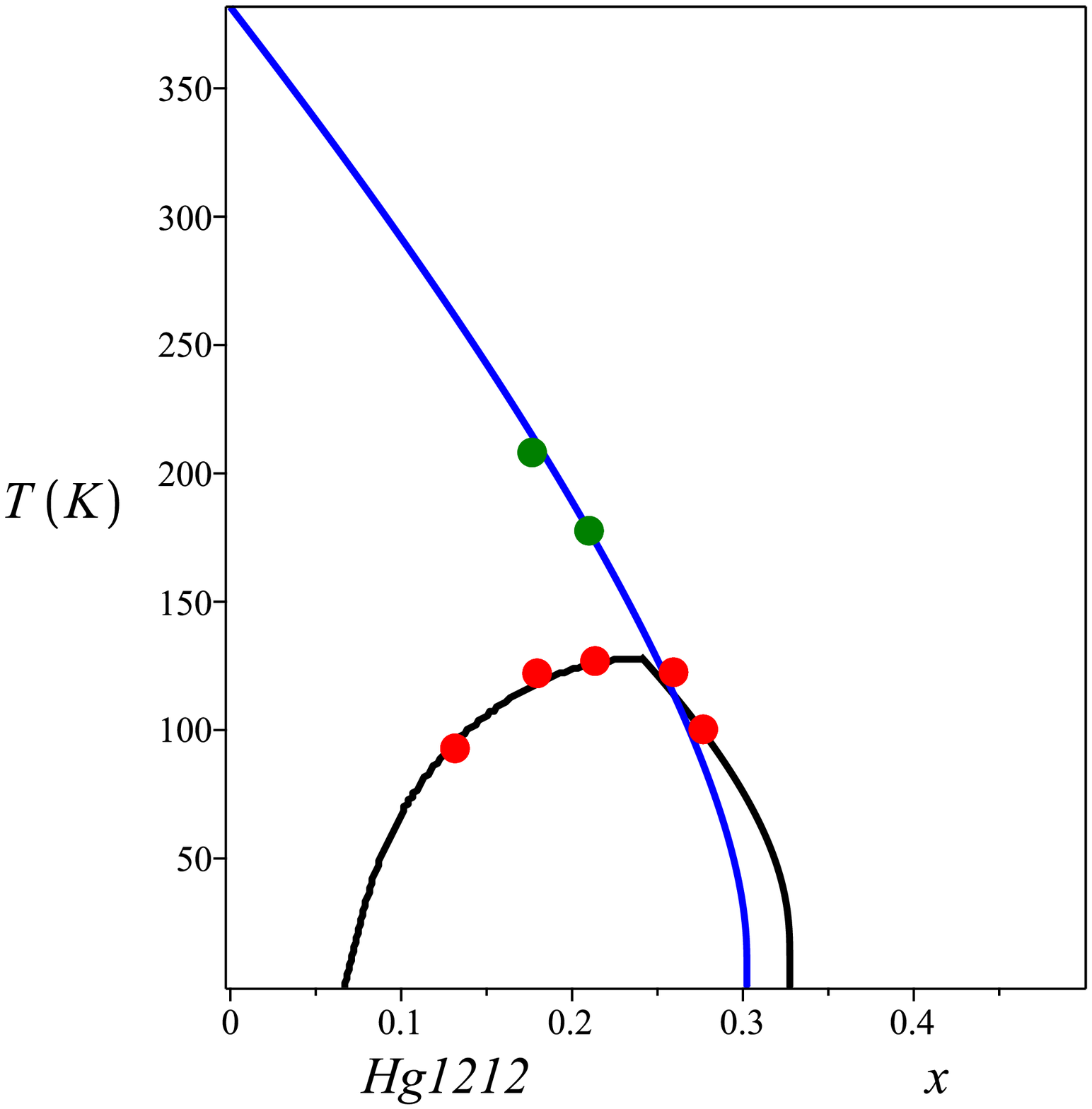}}
		\caption{Hg1212. Experimental data from \cite{02}}
	\label{f04}
\end{figure}
\begin{figure}
	[h!]
	\centerline{
		\includegraphics[scale=0.3]{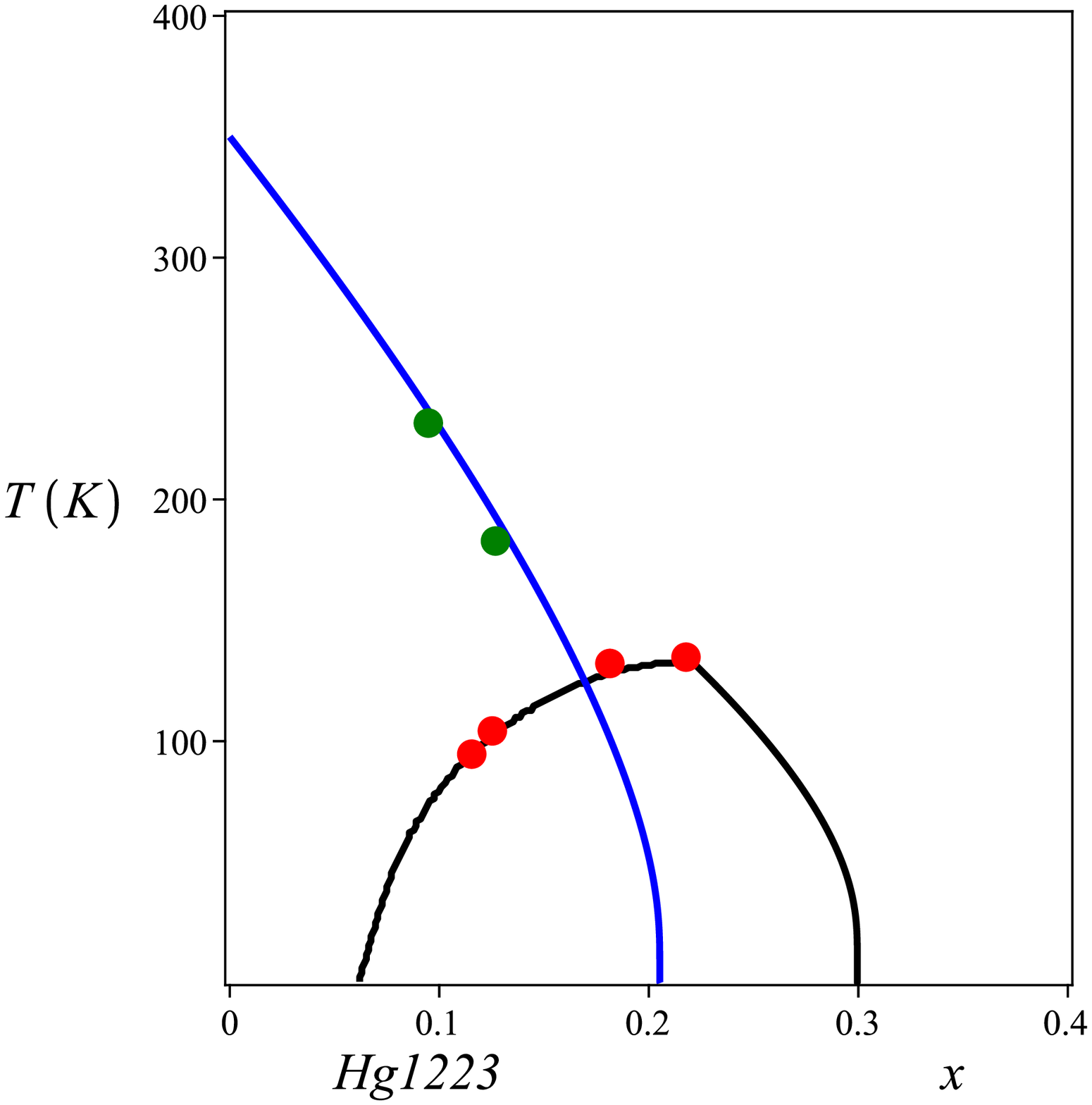}}
		\caption{Hg1223. Experimental
		data from \cite{02}}
	\label{f05}
\end{figure}
\begin{figure}
	[h!]
	\centerline{
		\includegraphics[scale=0.3]{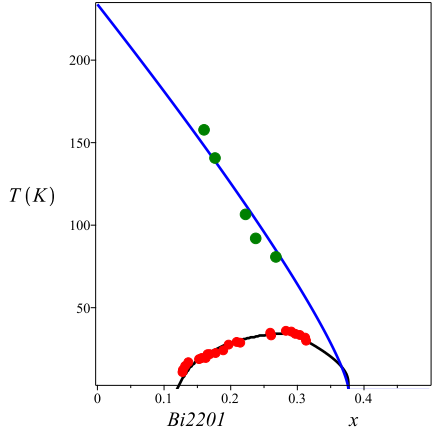}}
		\caption{Bi2201.Experimental data from \cite{03,04}}
	\label{f06}
\end{figure}
\begin{figure}
	[h!]
	\centerline{
		\includegraphics[scale=0.3]{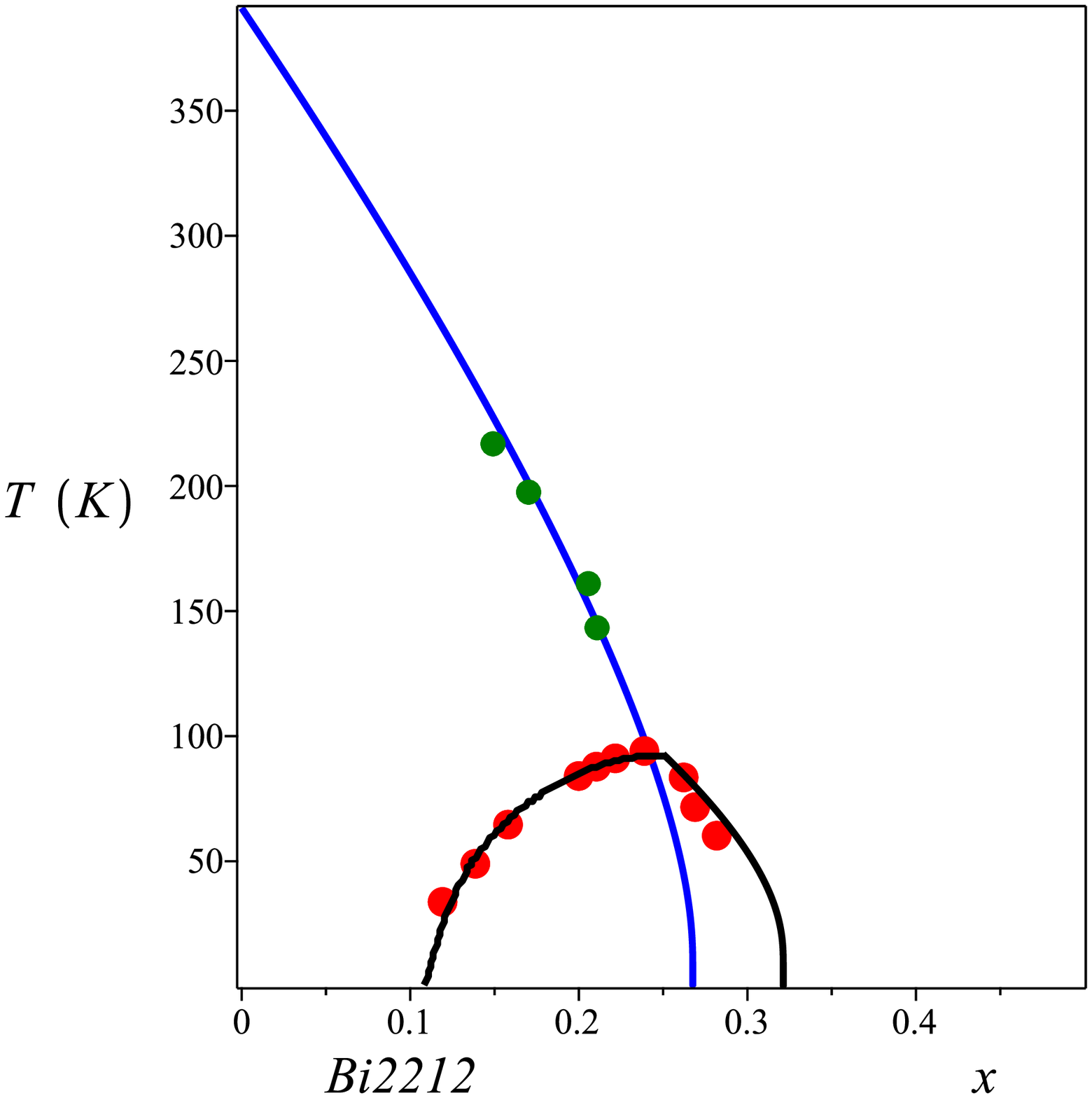}}
		\caption{Bi2212. Experimental data from \cite{05,06}}
	\label{f07}
\end{figure}
\begin{figure}
	[h!]
	\centerline{
		\includegraphics[scale=0.3]{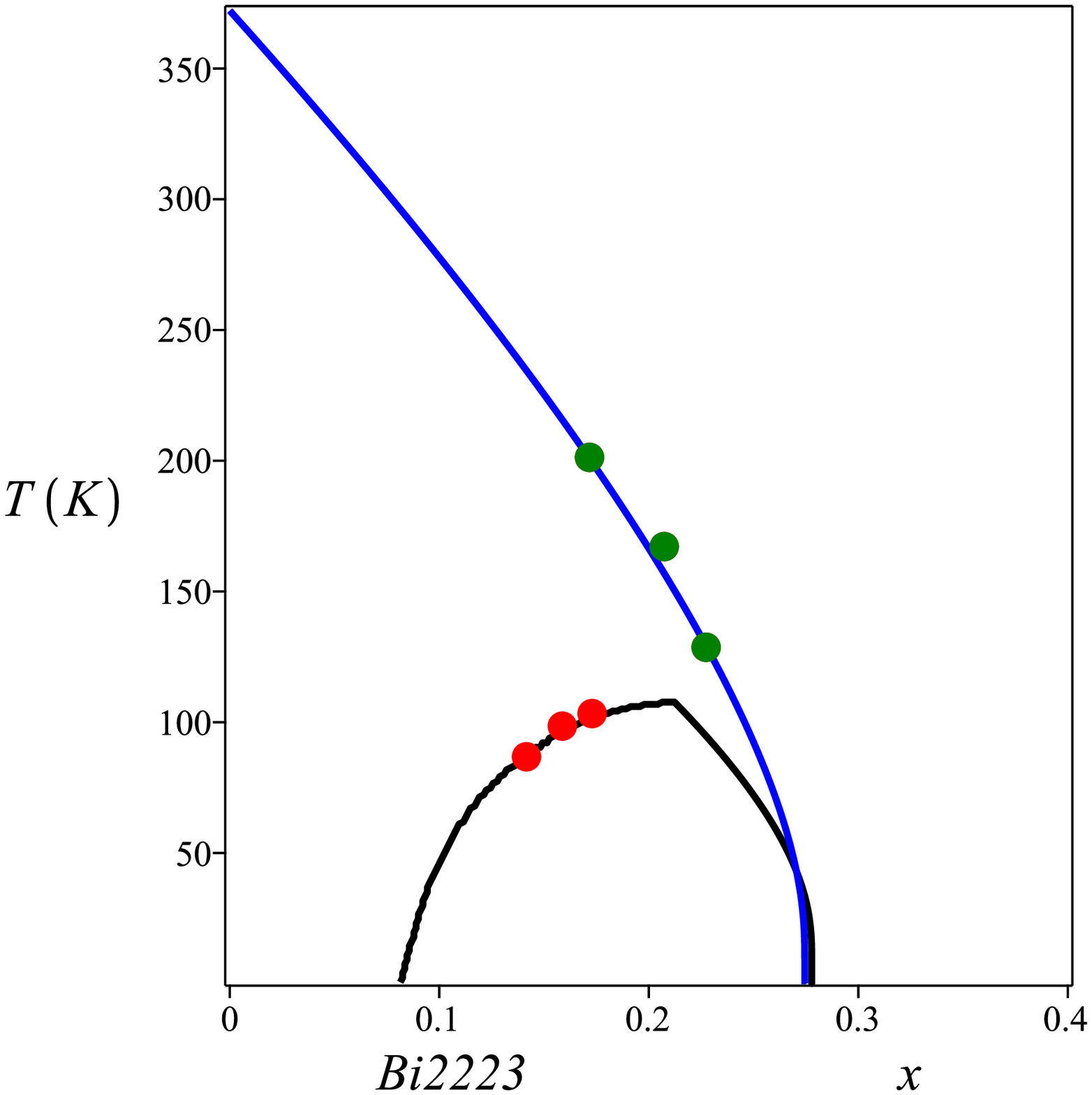}}
		\caption{Bi2223}. Experimental data from \cite{07}
	\label{f08}
\end{figure}
\begin{figure}
	[h!]
	\centerline{
		\includegraphics[scale=0.3]{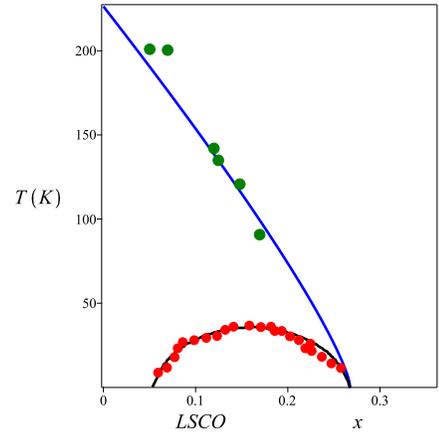}}
		\caption{LSCO. Experimental data from \cite{08,09}}
	\label{f09}
\end{figure}
\begin{widetext}

\begin{table}[!ht]
\begin{tabular}{|c|c|c|c|c|c|c|c|}
\hline\hline
           & N   & $T_{max} (eV)$ &  $x_0$   & $\gamma$ (eV) & $\eta(N)$    & $\eta(1)^{1/N}$ &  $\gamma x_0 \eta(1)^{1/N}$\\ \hline \hline
Bi2201 & 1       & 0.0030 &   0.29   &      0.012    & 0.23077      & 0.23077 & 0.0008030      \\ 
Bi2212  & 2      & 0.0080 &  0.25    &    0.0389     & 0.61538      & 0.48038 & 0.004671     \\
Bi2223   & 3     & 0.0093 &   0.212  &    0.049      & 0.74358      & 0.61337 & 0.00637    \\ \hline
Hg1201 & 1       & 0.0083 & 0.25     &    0.031      & 0.61577      & 0.61577 & 0.004772           \\
Hg1212 & 2       & 0.0111 &  0.24    &      0.044    & 0.80788      & 0.78471 &  0.00828          \\ 
Hg1223 & 3       & 0.0115 &  0.22   &     0.050     & 0.87192       & 0.85076 & 0.009357           \\ \hline
LSCO    & 1      & 0.0031 &   0.16  &   0.020       &   0.23870    & 0.23870 & 0.00076     \\ \hline
\end{tabular}
\caption{The parameters used for obtaining the $T_c(x)$ curves. $T_{max}$ and $x_0$ are fixed experimental inputs and $\eta$ is determined by the former. Only $\gamma$ has been adjusted, in order to fit the experimental data. The last column displays the value obtained for the combination $\gamma x_0 \eta(1)^{1/N}$.}
\label{t1}
\end{table}

\begin{table}[!ht]
\begin{tabular}{|c|c|c|c|c|c|c|}
\hline\hline
           & N   &   $\tilde{x}_0$   & $\tilde{\gamma}$ (eV) & $\tilde{\eta}(N)$    & $\tilde{\eta}(1)^{1/N}$ &  $\tilde{\gamma}\tilde{x}_0 \tilde{\eta}(1)^{1/N}$\\ \hline \hline
Bi2201 & 1       &    0.376   &      0.132    & 0.01618      & 0.01618 & 0.0008030       \\ 
Bi2212  & 2      &   0.24    &    0.153     &  0.50809             & 0.12720 & 0.004671     \\
Bi2223   & 3     &    0.245  &    0.1029      & 0.67205               & 0.25292 & 0.00637     \\ \hline
Hg1201 & 1       &  0.343     &    0.186      & 0.07480      & 0.07480 & 0.004772            \\
Hg1212 & 2       &   0.28    &      0.1082    & 0.53740              & 0.27349 &  0.00828           \\ 
Hg1223 & 3       &   0.18   &     0.1234     &   0.69159          & 0.42134 & 0.009357           \\ \hline
LSCO    & 1      &     0.269  &   0.180       &   0.01565    & 0.01565 & 0.00076       \\ \hline
\end{tabular}
\caption{The parameters used for obtaining the $T^*(x)$ curves. Only $\tilde\gamma$ has been adjusted in order to fit the experimental data. The last column displays the value obtained for the combination $\tilde{\gamma}\tilde{x}_0 \tilde{\eta}(1)^{1/N}$.}
\label{t2}
\end{table}
\end{widetext}
\begin{figure}
	[h!]
	\centerline{
		\includegraphics[scale=0.3]{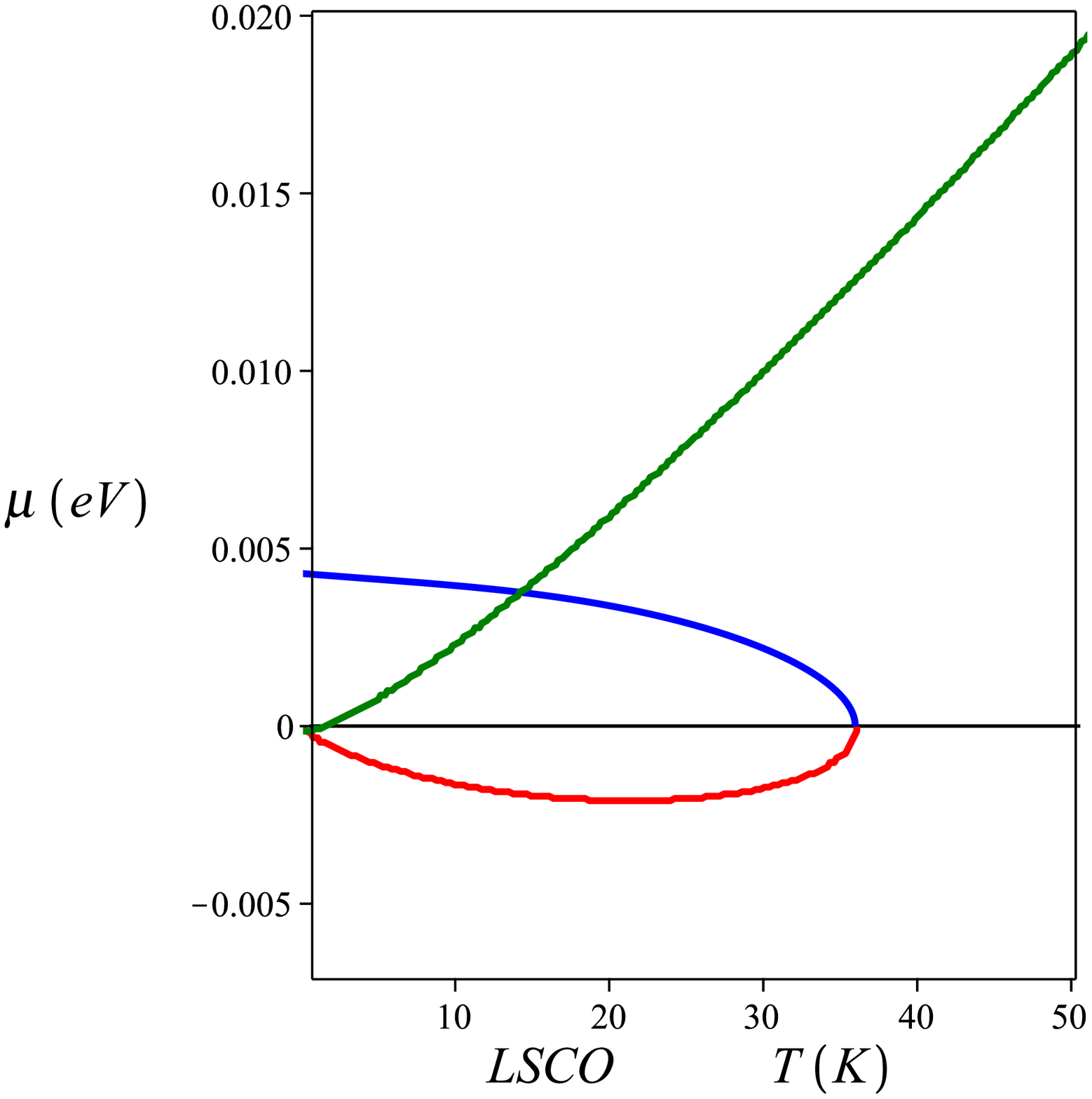}}
		\caption{The chemical potential for LSCO along the SC and PG transition curves: $T_c(x)$ (blue - underdoped and red - overdoped) and $T^*(x)$ (green)}
	\label{f09}
\end{figure}

\begin{figure}
	[h!]
	\centerline{
		\includegraphics[scale=0.3]{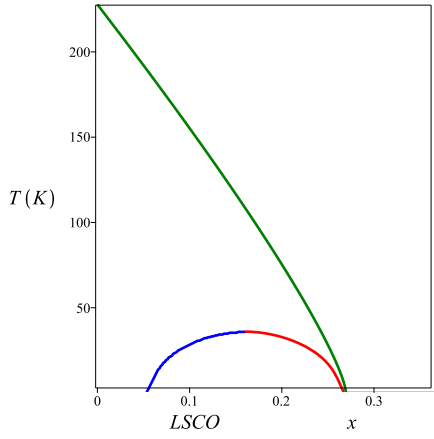}}
		\caption{The phase diagrams of LSCO, showing the transition curves $T_c(x)$ and $T^*(x)$.}
	\label{f09}
\end{figure}

\begin{figure}
	[h!]
	\centerline{
		\includegraphics[scale=0.3]{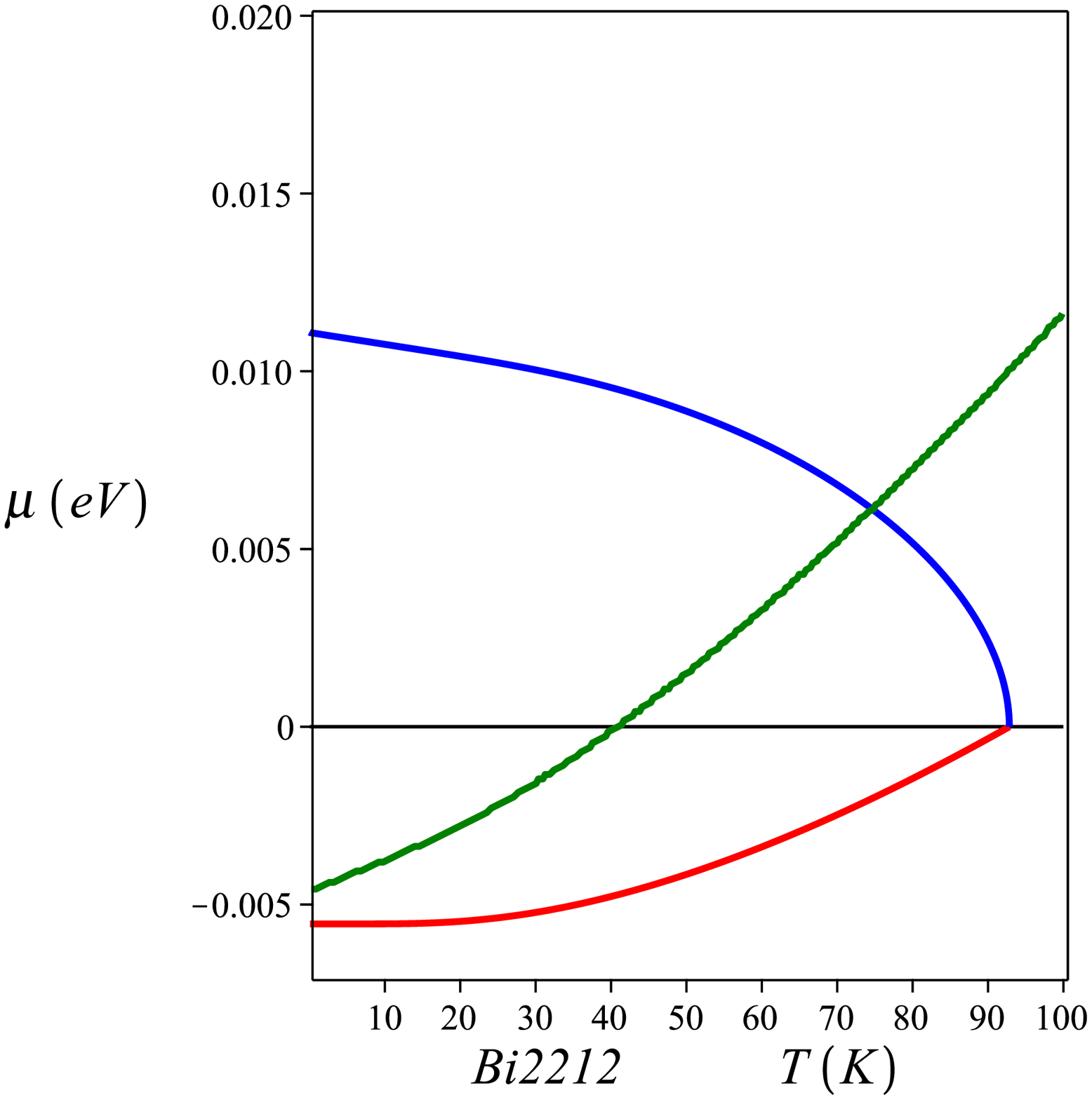}}
		\caption{The chemical potential for Bi2212 along the SC and PG transition curves: $T_c(x)$ (blue - underdoped and red - overdoped) and $T^*(x)$ (green)}
	\label{f09}
\end{figure}

\begin{figure}
	[h!]
	\centerline{
		\includegraphics[scale=0.3]{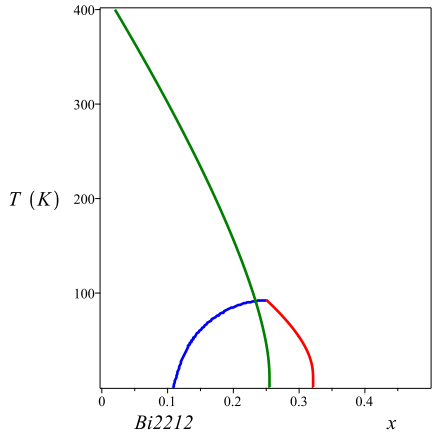}}
		\caption{The phase diagrams of Bi2212, showing the transition curves $T_c(x)$ and $T^*(x)$}
	\label{f09}
\end{figure}

\begin{figure}
	[h!]
	\centerline{
		\includegraphics[scale=0.3]{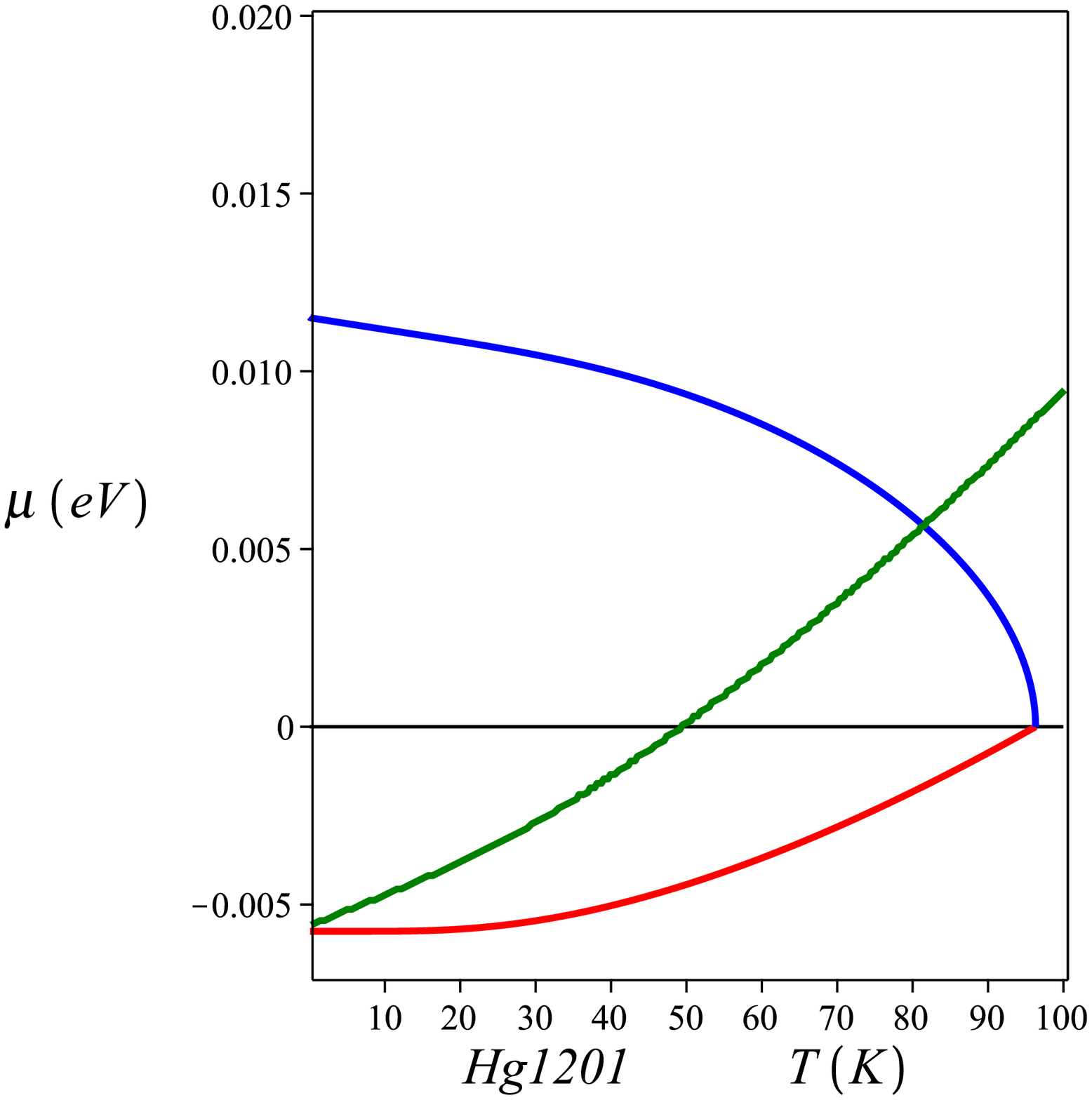}}
		\caption{The chemical potential for Hg1201 along the SC and PG transition curves: $T_c(x)$ (blue - underdoped and red - overdoped) and $T^*(x)$ (green)}
	\label{f09}
\end{figure}

\begin{figure}
	[h!]
	\centerline{
		\includegraphics[scale=0.3]{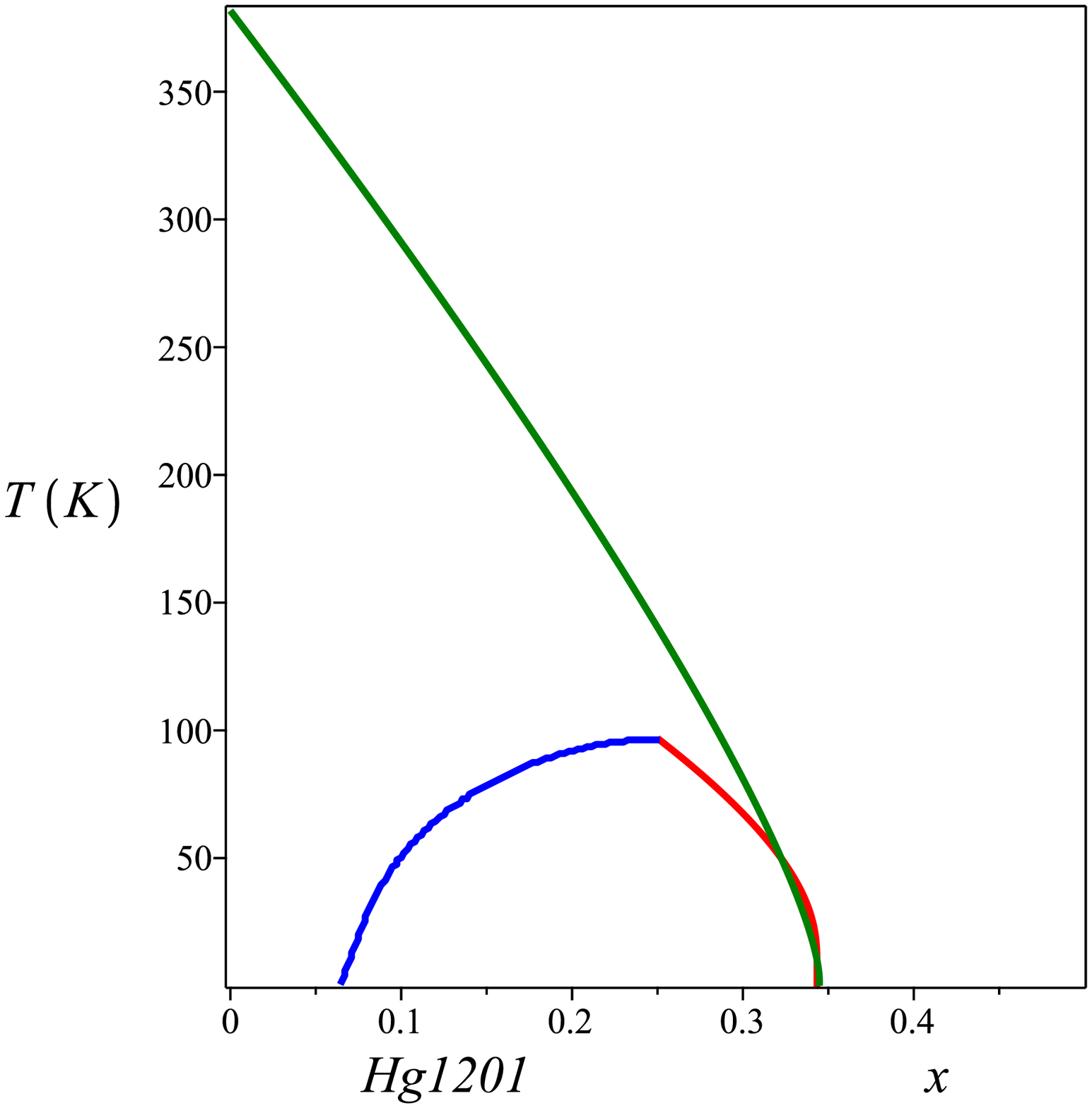}}
		\caption{The phase diagrams of Hg1201, showing the transition curves $T_c(x)$ and $T^*(x)$}
	\label{f09}
\end{figure}
{\bf 6) The Chemical Potential}
\bigskip

{\bf 6.1) $\mu(T)$ at the SC Transition} \\
\bigskip

We may invert Eq. (\ref{eqtc}), and thereby express the chemical potential as a function of the temperature on the $T_c(x)$-curve, namely

\begin{eqnarray}
	\begin{cases}
			\mu(T_c) = - T_c \ln\left[2^{T_{max}/T_c}-1 \right],\hspace{0.5cm}x > x_{0}\\
		\mu(T_c)=2\ln2 \left( T_{max}-T_c\right )+T_c\left[1- e^{-\mu/T_c} \right]  ,\hspace{0.3cm} x < x_{0}\\
			T_{c}(x) =\frac{\ln2 \,\ T_{max}}{\ln2 + \frac{|\mu_0(x)|}{2T_{c}(x)} - \frac{1}{2}\left(1-e^{-\frac{|\mu_0(x)|}{T_{c}(x)}}\right)},\hspace{0.3cm} \text{LSCO},\forall x
			\label{eqmu}
		\end{cases}
\end{eqnarray}
where the last two equations are implicit for the chemical potential $\mu(T)$.

In the three next figures, we plot the solution $\mu(T)$ for the above equations and compare with the corresponding figures for the phase diagram $T\times x$. Segments with a given color correspond to the corresponding segment in the phase diagram graph.

Observe that what determines whether the PG transition curve ends inside the SC dome or not are the values of the chemical potential at $T=0$ for the SC and PG curves, namely $\mu(T=0)$ and $\tilde\mu(T=0)$. In the examples considered here, such values coincide for LSCO and Hg1201 and differ for Bi2212.

\bigskip
\vfill
\eject
{\bf 7) Applied Pressure Effects }\\
\bigskip

{\bf 7.1) Pressure Influence on the SC coupling $g_S$}\\
\bigskip
We have seen in \cite{M1} that, under an applied pressure $P$, the SC coupling varies as
\begin{equation}
 g_S(P)= g_S e^{P/\kappa} ,  
 \label{gsp}   
\end{equation}
where $\kappa$ must be adjusted.

The function $\eta(g_S)$, consequently, changes as
\begin{equation}
    \eta(P)=1-\frac{g_c}{Ng_S(P)}
    \label{eta1}
\end{equation}

The SC transition temperature, $T_c(x;P)$, by its turn, will be modified as
\begin{widetext}
\begin{equation}
			T_{c}(x;P) =\frac{\frac{\Lambda \eta(P)}{2}}{\ln2 + \frac{\gamma(P)x_0(P)\left [1-\frac{x}{x_0(P)}\right]}{T_{c}(x;P)} - \frac{1}{2}\left[1-e^{-\frac{\gamma(P)x_0(P)\left[1-\frac{x}{x_0(P)}\right]}{T_{c}(x;P)}}\right]},
			\label{eqtc1}
	\end{equation}
	\end{widetext}
Now, from (\ref{m33}), it follows that
\begin{equation}
    \gamma(P) x_0(P) =\gamma(0) x_0(0) \frac{\eta(N=1;0)^{1/N}}{\eta(N=1;P)^{1/N}},
    \label{m6}
\end{equation}
Writing
\begin{equation}
    \gamma(P)  =\gamma(0) f(P)\ \ ;\ \ x_0(P) = x_0(0) g(P)
    \label{m7}
\end{equation}
then, it follows that
\begin{equation}
  f(P)g(P)=\frac{\eta(N=1;0)^{1/N}}{\eta(N=1;P)^{1/N}}\equiv A(P)
  \label{m8}
\end{equation}
 Choosing $g(P)= A^{1/2}(P)$, we have $f(P)= A^{1 /2}(P)$.
Inserting in (\ref{m6}) and solving for $T_c(P)$, we obtain, for Hg1212 and Hg1223, using the parameters of Table 1 and adjusting only $\kappa= 9 \ GPa$ for the former and $\kappa= 4\  GPa $ for the latter:
\begin{figure}
	[h!]
	\centerline{
		\includegraphics[scale=0.3]{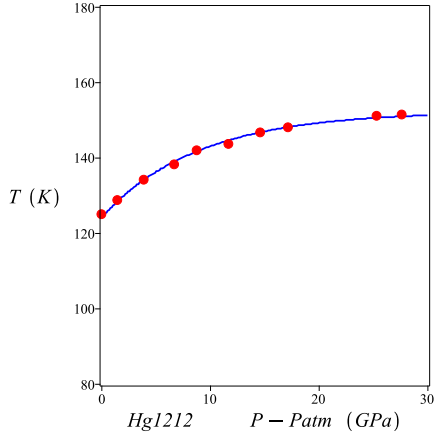}}
		\caption{$T_c(x=x_0(0);P)$ for Hg1212. The solid line is given by our analytical expression (\ref{eqtc1}) by adjusting a single parameter, namely, $\kappa= 9 \ GPa $. The experimental data are from \cite{HgP} }
	\label{f009}
\end{figure}
\begin{figure}
	[h!]
	\centerline{
		\includegraphics[scale=0.3]{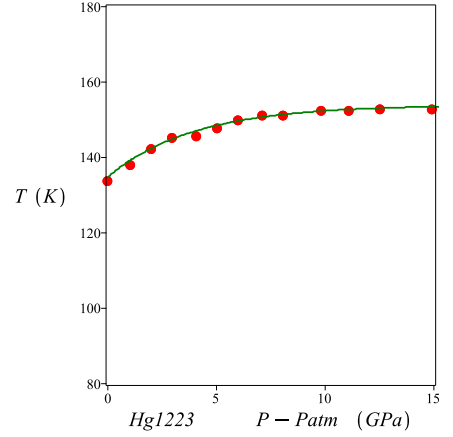}}
		\caption{$T_c(x=x_0(0);P)$ for Hg1223. The solid line is given by our analytical expression (\ref{eqtc1}) by adjusting a single parameter, namely, $\kappa= 4\  GPa $. The experimental data are from \cite{HgP}}
	\label{fa9}
\end{figure}
     
\vfill
\eject

\bigskip
{\bf 8) Conclusion}\\

The study we report here brings forth, among other things, the completion of the solution of two fundamental issues in the physics of high-Tc cuprate superconductors, which we started in \cite{M1} and continued in \cite{M2,M3}.

The first one concerns the issue as to what is the ultimate nature of the interaction that makes the holes doped into the oxygen $p_x$ and $p_y$ orbitals to form Cooper pairs and, thereby, to exhibit a SC phase.

The second one concerns the existence of a PG order parameter, that would be different from zero in the whole PG phase and zero outside of it, in such a way that it could be used to unequivocally characterize such a phase. 

We completely elucidate these two points, providing the final pending details, which are required for the full comprehension thereof.

 We have shown that the interaction that leads to Cooper pair formation derives from the Kondo-like magnetic interaction that exists between the itinerant oxygen holes and the localized copper spins. A crucial issue, however, is the peculiar arrangement of the oxygen p-orbitals depicted in Fig. 2, which guarantees an attractive interaction for {\it all} nearest neighbor holes.
 
 Below Tc, the energetically most favorable configuration of the oxygen p-orbitals, which hybridize with the copper ions d-orbitals is the one represented in red-white, in Figs. 2 and 6. Notice that this configuration breaks the 90$^\circ$ rotation symmetry and naturally leads to DDW SC and PG order parameters.  \cite{ddw,ddw0,ddw1}.  Besides that, remarkably, this orbital arrangement leads to an ever attractive effective interaction between neighboring holes, by means of a mutual magnetic interaction with the closest copper ion.

Observe that, comparing the configurations that repeat themselves, given in Fig. 1, with the one in Fig. 3, we conclude that in order to produce the orbital arrangement that will produce the attractive interaction, which leads to superconductivity, the system undergoes a dimerization that resembles the one found in polyacetylene \cite{ssh}. Furthermore, the dimer configuration exhibits an AF ordering of the copper spins, associated with the super-exchange mechanism.

Secondly, our analysis shows that the spectral density which results from an energy dispersion relation containing a d-wave gap parameter, such as the one in (27), exhibits two sharp peaks, symmetrically placed around the Fermi level at  positions proportional to $\pm \frac{M^2}{V^2 +M^2}$. In between the two peaks, the density of states is depleted but non-vanishing, that is why it is properly called ``pseudogap''. Should the gap parameter be uniform in $\mathbf{k}$-space, we would have a real gap, with the spectral density actually vanishing in between the peaks.
The peak separation in the spectral density, which is observed along the whole PG region is the PG order parameter. We have seen that it derives directly from the 90$^\circ$ rotation symmetry breakdown that leads to expression (27) for the pseudogap.

We calculate the spectral density corresponding to the dispersion relation in the presence of a d-wave symmetric pseudogap order parameter, given by (37) and (38) and show that it leads to a doubled peaked spectral density, such that the peaks, symmetrically placed around the Fermi level delimit a region where the spectral density is depleted. Our theoretical results are in good agreement with the experimental data for Bi2212 \cite{sw}.\\
\bigskip
{\bf  Acknowledgements}\\

The author is grateful to Nigel Hussey for interesting and stimulating conversations. This study received partial financial support from CNPq, FAPERJ and CAPES.
\bigskip

\vfill
\eject

\end{document}